\documentstyle[12pt,preprint,prd,aps,epsf]{revtex}
\tightenlines
\begin{document}
\draft
\input psfig
\begin{titlepage}

\title{\bf Where do we stand with solar neutrino oscillations?}

\author{J. N. Bahcall}
\address{School of Natural Sciences, Institute for Advanced
Study\\
Princeton, NJ 08540\\}
\author{P. I. Krastev}
\address{University of Wisconsin, Madison, WI 53706}
\author{A. Yu. Smirnov}
\address{ICTP, Trieste, Italy}

\maketitle
\begin{abstract}
We determine parameters 
for MSW and vacuum oscillations (active and sterile neutrinos) 
that are allowed by  separate, and  collective,  imposition  of the
constraints from total event rates in the chlorine, GALLEX, SAGE,
and SuperKamiokande experiments (504 days), 
the SuperKamiokande  
energy spectrum, and the SuperKamiokande zenith-angle dependence.
The  small
mixing angle MSW solution is  acceptable at  $7$\% C.L. ($8$\%
for sterile $\nu$'s) and the  vacuum
solution is acceptable at  $6$\% C.L. .
The best-fit global MSW solution for active neutrinos is: $\Delta m^2  =  
5\times 10^{-6} {\rm eV}^2 ,
\sin^22\theta  = 5.5\times 10^{-3}~ $ (and for sterile neutrinos:
$\Delta m^2  =  
4 \times 10^{-6} {\rm eV}^2 , \sin^22\theta  = 7\times 10^{-3}~ $).
For vacuum oscillations, the best-fit solution is:
\hbox{$\Delta m^2  =  
6.5 \times 10^{-11} {\rm eV}^2 ,
\sin^22\theta  = 0.75~ $ .}
An arbitrary combination of undistorted (no oscillations) 
$pp$, ${\rm ^7Be}$, ${\rm ^8B}$, and CNO neutrino
fluxes is inconsistent with the combined data sets at the $3.5\sigma$
C.L., independent of astrophysical considerations.
We use improved calculations of solar model
fluxes, neutrino absorption cross sections and energy spectra, and a detailed
evaluation of regeneration effects.

\end{abstract}
\end{titlepage}
\newpage

\section{INTRODUCTION}
\label{intro}

It is now 30 years since the first recognition of the solar neutrino
problem \cite{davis68,bahcall68,bahcall76}. 
In the first two decades of solar neutrino research~\cite{bahcall89}, the 
problem consisted only of the discrepancy between theoretical calculations 
based upon on a standard solar model (with the implicit assumption
that neutrinos created in the solar interior reach the earth unchanged)
and the observations of the capture rate in the chlorine solar neutrino
experiment. 

In recent years, four new experiments
(Kamiokande \cite{kamiokande}, GALLEX \cite{GALLEX}, SAGE \cite{SAGE}, 
and SuperKamiokande \cite{superkamiokande300,superkamiokande374,superkamiokande504})
have reported results.
All four experiments confirm the original detection of solar neutrinos
with lower neutrino fluxes than predicted by standard solar models. In
addition, the Kamiokande and SuperKamiokande experiments demonstrate
directly that the neutrinos come from the sun by showing that recoil
electrons are scattered in the direction along the sun-earth axis.

In April 1996, the SuperKamiokande experiment initiated a new era of
high-precision and high-statistics solar neutrino research. 
The first 504 days of data from
SuperKamiokande \cite{superkamiokande504}, 
when combined with data from earlier experiments on
solar neutrinos,  provide important constraints on 
the MSW \cite{ref:msw} 
 and vacuum oscillation \cite{ref:vac}
solutions of the
solar neutrino problem. 

The definitive analysis  of the implications of the 
SuperKamiokande data set must await the comprehensive Monte Carlo
study that can only be performed by the SuperKamiokande
Collaboration. However, 
the excellent agreement of the results from the first 300 days of
SuperKamiokande operation \cite{superkamiokande300}
with the results obtained after 374 days \cite{superkamiokande374} and
after 504 days \cite{superkamiokande504} 
shows the robustness of the results.  Therefore, 
with more than a year's worth of data available, 
this is an appropriate time to take stock of what has been achieved
and what further challenges lie ahead. 
The analysis presented here 
will, we hope, 
 be useful in guiding experimental plans for other detectors such
as SNO~\cite{McD94} and BOREXINO~\cite{borexino}.  
We also hope that our summaries  of the analysis
techniques and the  theoretical input 
data (see especially Sec.~\ref{technique} 
and the Appendix), as well as the indicated
results, will be helpful to others  who will make similar
studies.
The results from different theoretical analyses 
should be compared  with each other, and with the comprehensive 
studies by the SuperKamiokande Collaboration using their detailed
Monte Carlo simulation, in order to test the robustness
of the inferences about neutrino parameters.

In this paper, we explore the implications of the $504$ day  data set
from SuperKamiokande together with the results from the chlorine,
GALLEX, and SAGE experiments. 
For a concise summary of our conclusions, the reader is advised to
skip directly to Sec.~\ref{discussion} and then to return to this
introduction. 

We use improved neutrino interaction cross 
sections~\cite{bahcallga97,bahcalletal96,bahcall88,sirlin}, the most
accurate neutrino spectra~\cite{bahcallga97,balisi}, and 
the results of a recent reevaluation, the BP98 model, 
by Bahcall and Pinsonneault
of the standard model
neutrino fluxes \cite{BP98}.  The BP98 model is   based upon a comprehensive 
examination of all of the
available nuclear fusion  data \cite{adelberger98}
that was carried out under the auspices of the 
Institute of Nuclear Theory (INT).
For solar neutrino research, the most important nuclear
physics parameter is the low energy cross section factor, $S_{17}$,
for the reaction ${\rm ^7Be}(p,\gamma){\rm ^8B}$, which gives rise to the critical 
${\rm ^8B}$ neutrinos.
The INT  normalization, $S_{17}({\rm INT}) = 19^{+4}_{-2}$~eV~b
\cite{adelberger98},	is about $1\sigma$ less than the 
 previously standard Caltech (CIT) cross
section factor \cite{johnson92}, $S_{17}({\rm CIT}) = (22.4 \pm 2.1)$~eV~b.
The CIT normalization was
computed by taking the weighted average over all the published
experimental data while the INT normalization was computed by
including only the results from the two most recent and best
documented experiments.
Because the uncertainty in  the ${\rm ^8B}$ production cross section is the
most important  uncertainty in predicting solar neutrino fluxes\cite{BP98}, 
we present results in this paper for 
both the INT and the CIT normalizations of $S_{17}$.

We also present
calculations for oscillations into sterile neutrinos as well as into
the more familiar active neutrinos.  In what follows, we shall always
mean oscillations into active neutrinos unless we explicitly include
the adjective `sterile.'
We do not consider here the intermediate case of  oscillations partly
into active neutrinos and partly   into sterile neutrinos,
although this is a logical possibility.
For plausible assumptions 
(see e.g. ~\cite{giunti}), consistent with the results of the  CHOOZ
experiment~\cite{chooz}, the 
oscillations of solar neutrinos are well described by the
two-generation formalism. Therefore, we consider explicitly here only
two neutrino generations.

We begin by considering in Sec.~\ref{fuss} how well (or rather how 
poorly) the
results of solar neutrino experiments are described by the combined
predictions of the standard solar model and the minimal electroweak
theory (which implies that nothing happens to the neutrinos after they
are created). In Sec.~\ref{technique} we briefly summarize the
ingredients and techniques used in our analysis and give references to
the original sources for the improved neutrino flux calculations,
the associated uncertainties in the fluxes, the improved neutrino
energy spectra and 
interaction cross sections, the methods including theoretical errors,
and the techniques for carrying out the theoretical calculations.
Details of the statistical analysis are provided in the Appendix.
We determine  in Sec.~\ref{averagerates} 
the regions that are allowed in neutrino
parameter space for MSW and vacuum oscillations to either active
or sterile neutrinos provided only the total event rates in the
neutrino experiments are considered. 
In Sec.~\ref{zenith} we determine  the 
implications of the zenith-angle dependence of the
SuperKamiokande event rates.  We quantify in Sec.~\ref{spectralshape}
the distortion of the recoil electron energy spectrum measured by
SuperKamiokande, determining the slope parameter and the excluded
regions of oscillation parameters. 
We impose in Sec.~\ref{global}
all of the constraints, total rates, electron recoil energy
spectrum, and zenith-angle dependence, in global fits and determine
the range of oscillation parameters that are consistent with all the
data.  
We present in Sec.~\ref{discussion} our summary and 
overview of where we stand in understand the discrepancies between
standard model predictions and the results of solar neutrino
experiments and in 
the determination of neutrino parameters
from solar neutrino experiments.

\section{WHAT IS ALL THE FUSS ABOUT?}
\label{fuss}

Why are so many papers being written about non-standard physics
implied by solar neutrino experiments?  This section provides two
answers to this question. In Sec.~\ref{ss:ssms}, we show that all of the 
19 standard solar model calculations published in refereed journals
in the last $10$ years predict neutrino fluxes that are 
in reasonable agreement with each other. In Sec.~\ref{ss:mit}, we show
that the measured  rates and their uncertainties 
in solar neutrino experiments are inconsistent
with any  combination of the solar neutrino fluxes that does not
include a spectrum distortion--which requires physics beyond the
standard electroweak model.  We summarize in Sec.~\ref{ss:discussionsm} the
comparisons between the predictions of the standard model--minimal
electroweak theory plus standard solar model--and the results of 
solar neutrino experiments.

\subsection{The last decade of standard solar models}
\label{ss:ssms}

Figure~\ref{fig:independent}
displays the 
calculated  ${\rm ^7Be}$ and ${\rm
^8B}$ neutrino fluxes 
for all $19$ standard solar models with which we are
familiar which have been published in the last $10$ years in refereed
science journals.  
The fluxes are normalized by 
dividing each
published value by the flux from the BP98
solar model~\cite{BP98}; the abscissa is the
normalized ${\rm ^8B}$ flux and the ordinate
 is the normalized ${\rm ^7Be}$
neutrino flux.  The rectangular box shows the estimated 
$3\sigma$ uncertainties in the
predictions of the BP98 solar model.
The abbreviations, which  indicate references to individual models, are
identified in the caption of Figure~\ref{fig:independent}.

All of the solar model results from different groups 
fall within the estimated 3$\sigma$
uncertainties in the model predictions (with the exception of the
Dar-Shaviv model whose results have not been reproduced by other groups).
This agreement 
demonstrates the robustness of the predictions since the
calculations use different computer codes (which achieve varying
degrees of precision) and 
involve a variety of choices for the nuclear
parameters, the equation of state, the stellar radiative opacity, 
the initial heavy element abundances, and the physical processes
that are included.

The largest contributions to  the dispersion in values in
Figure~\ref{fig:independent} are due to the
choice of the normalization for  
$S_{17}$ (the production
cross-section factor for ${\rm ^8B}$ neutrinos) and the
inclusion, or non-inclusion, of element diffusion in the 
stellar evolution codes.
The effect in the plane of Fig.~\ref{fig:independent} 
of the normalization of
$S_{17}$ is shown by the difference between the point for BP98
(1.0,1.0), which was computed using the INT normalization, and the
point at (1.18,1.0) which corresponds to the BP98 result with the CIT
normalization.   

Helioseismological observations have shown recently~\cite{reliable}
that diffusion is occurring
and must be included in solar models, so that the most recent models
shown in Fig.~\ref{fig:independent} now all include helium and heavy
element diffusion.  By comparing a large number of earlier models,
it was shown that 
all published standard  solar
models give the same results for solar neutrino fluxes to an accuracy
of better than 10\% if the same input parameters and physical
processes are included~\cite{BP92,BP95}.

How do the 
observations from the solar neutrino
experiments agree with the solar model calculation?  

Table~\ref{datarates} summarizes the solar neutrino experimental rates
that have been measured in the five  experiments.
We have compared the observed rates 
with the calculated, standard model values, combining  quadratically
the theoretical solar model and experimental
uncertainties, as well as the uncertainties in the neutrino cross
sections.
Since the GALLEX and SAGE experiments measure the same quantity, we
treat the weighted average rate in gallium as one experimental number.
We adopt the SuperKamiokande measurement as the most precise direct
determination of the higher-energy ${\rm ^8B}$ neutrino flux.

Using the predicted fluxes from the BP98
model, the $\chi^2$ for the fit to the three  experimental rates
(chlorine, gallium, and SuperKamiokande) is

\begin{equation}
\chi^2_{\rm SSM} \hbox{(3 experimental rates)} = 61\ .
\label{chitofit}
\end{equation}
The result given in Eq.~(\ref{chitofit}), which is approximately 
equivalent to  a
$20\sigma$ discrepancy,  is a quantitative expression 
of the fact that the standard model predictions do not fit the
observed solar neutrino measurements.

\subsection{Model-independent tests}
\label{ss:mit}

Suppose 
(following the precepts of Hata {\it et al.}~\cite{hata94}, Parke~\cite{parke95},
 and Heeger and Robertson~\cite{robertson})
we now ignore everything we  have learned  about  solar models over
the last $35$ years and allow
the important $pp$,
${\rm ^7Be}$, and ${\rm ^8B}$ fluxes to take on any non-negative
values.  What is the minimum value of $\chi^2$ for the $3$ experiments,
when the only constraint on the fluxes is the requirement that the
luminosity of the sun be supplied by nuclear fusion reactions among
light elements?  We include the nuclear physics inequalities between
neutrino fluxes (see section 4 of Ref.~\cite{howwell}) that are
associated with the luminosity constraint and maintain the standard
value for the essentially 
model-independent  ratio of $pep$ to $pp$ neutrino fluxes.

\subsubsection{With SSM value for CNO neutrinos}
\label{sss:inclcno}

We begin by allowing the $pp$, ${\rm ^7Be}$, and ${\rm ^8B}$ neutrino fluxes to
be arbitrary parameters (subject to the luminosity constraint), but 
constrain the $^{13}$N and $^{15}$O fluxes to be equal to the values
predicted by the standard solar model~\cite{BP98}.
There is therefore one degree of freedom for three experiments plus
the luminosity constraint and three freely chosen neutrino fluxes.
The best fit for arbitrary $pp$, ${\rm ^7Be}$, and ${\rm ^8B}$  fluxes 
is obtained for ${\rm ^7Be/(^7Be)}_{\rm SSM} = 0.0$,
${\rm ^8B/(^8B)}_{\rm SSM} = 0.44$, and $pp = 1.08$, where

\begin{equation}
\chi^2_{\rm minimum} \hbox{(3 experimental rates;
~arbitrary $pp$, ${\rm ^7Be}$, ${\rm ^8B}$)} = 24.1\ . 
\label{chiall}
\end{equation}
The best-fit solution
differ considerably from the measured
values for the radiochemical experiments, with 
$3.4$ SNU for chlorine and $95.5$ SNU for gallium, but is in good
agreement with the measured value ($0.474$ of BP98) for SuperKamiokande
 (cf. Table~\ref{datarates}). 

There are no acceptable fits at  a C.L. of more  than $99.99$\%
($4\sigma$ result). 

\subsubsection{CNO neutrinos assumed missing}
\label{sss:nocno}

The fit can be 
improved if we set the CNO neutrino fluxes equal to zero.
\footnote{If the CNO neutrino fluxes are allowed to vary as free
parameters, the minimum $\chi^2$ is achieved for zero CNO fluxes.}
Then, the same search for arbitrary $pp$, ${\rm ^7Be}$, and
${\rm ^8B}$ neutrino fluxes leads to a 
 best fit with ${\rm ^7Be/(^7Be)}_{\rm SSM} = 0.0$,
${\rm ^8B/(^8B)}_{\rm SSM} = 0.46$, and $pp/(pp)_{\rm SSM} = 1.10$ ..
The minimum value of $\chi^2$ is

\begin{equation}
\chi^2_{\rm minimum} \hbox{(3 experimental data;
~arbitrary $pp$, ${\rm ^7Be}$, ${\rm ^8B}$; CNO = 0)} = 7.3\ .
\label{chinoCNO}
\end{equation}
The best-fit solution
has $2.9$ SNU for chlorine and $85.3$ SNU for gallium, 
and $0.46$ of BP98 for SuperKamiokande
 (cf. Table~\ref{datarates}). 
Although these three best-fit values are not far from the three
measured values, the best-fit values were found by an extensive
 computer search
with three free parameters and only one physical constraint, that the
nuclear energy generation rate correspond to  the total
luminosity of the sun.

There are no acceptable solutions at the $99$\% C. L. ($\sim 3\sigma$
result). 

Figure~\ref{fig:independent} shows the best-fit solution and the
 $1\sigma $
 --$3\sigma $ contours.  The $1\sigma$ and $3\sigma$ limits  
were obtained by requiring
that $\chi^2 =  \chi^2_{\rm min} + \delta\chi^2 $, where for $1\sigma$
$\delta\chi^2 = 1 $ and for $3\sigma$ $\delta\chi^2 = 9$.  All of
 the standard 
model solutions lie far from the best-fit solution and  even lie far 
from the $3\sigma$ contour.

\subsection{Discussion of comparisons with standard model}
\label{ss:discussionsm}

The searches for best-fit solutions that are 
described by Eq.~({\ref{chiall}) and Eq.~(\ref{chinoCNO})
have three independent experimental data points and $1$ additional
constraint (the luminosity constraint) with three free parameters
(the $pp$, ${\rm ^7Be}$, and ${\rm ^8B}$ neutrino fluxes).  
We conclude from  Eq.~({\ref{chiall}) and Eq.~(\ref{chinoCNO}) that
all solutions with undistorted energy spectra are ruled out at the
$99$\% C.L. (with the CNO neutrino fluxes set equal to zero, see 
Eq.~(\ref{chinoCNO}) or at more than $99.9$\% C.L. (with the CNO fluxes
equal to the values predicted by the standard solar model).

The results in this section (see Eq.~\ref{chitofit} and 
Fig.~\ref{fig:independent}) 
show that the combined standard model, the 
standard solar model plus minimal standard electroweak theory, provides
a bad fit to the observed rates in solar neutrino experiments.
Moreover, Eq.~(\ref{chiall}) and Eq.~(\ref{chinoCNO}) 
show that one cannot get a
good fit to the observed rates using any  combination of
undistorted neutrino energy spectra. All so-called astrophysical
solutions give a poor description of the observed experimental rates.
This is what all the fuss is about.

If we drop the physical requirement that the fluxes be positive
definite, the minimum $\chi^2$ ($\chi^2_{\rm min} = 0.1$)
occurs--quite remarkably--  
 for a negative value of the ${\rm ^7Be}$
flux; $\phi({\rm ^7Be})_{\rm min} = -0.45 \times \phi({\rm ^7Be})_{\rm SSM}$. 
This unphysical 
result is a reflection of what has become  known as the problem of 
`` the  missing ${\rm
^7Be}$ solar neutrinos.''.  One reason that the ${\rm ^7Be}$ neutrinos appear
to be missing (or have a negative flux) is that the two gallium
experiments, GALLEX and SAGE, have an average event rate of $72.3 \pm 5.6$
SNU,
which is  fully accounted for in the standard solar model 
by the fundamental $pp$ and $pep$ 
neutrinos ($72.4$ SNU)~\cite{BP92,BP95,BP98}. 
In addition,
the ${\rm ^8B}$ neutrinos
that are observed in the Kamiokande and SuperKamiokande 
experiments will produce about $6$
SNU in the gallium experiments, unless new particle physics changes
the neutrino energy spectrum.
A second reason that the ${\rm ^7Be}$ flux appears to be missing is that 
 (see~\cite{bethe90}) the  SuperKamiokande ${\rm ^8B}$ neutrino
flux alone corresponds to $2.70(1 \pm 0.11)$ SNU in a chlorine
detector (combining quadratically the theoretical cross section errors
with the SuperKamiokande measurement errors), which is to be
compared to the total observed
rate in the Homestake chlorine experiment of $(2.56 \pm 0.23)$ SNU.
The observed rate includes neutrinos from ${\rm ^8B}$, ${\rm ^7Be}$, 
CNO, and pep. 
Obviously, there is no room in the measured chlorine rate for a
significant ${\rm ^7Be}$ neutrino contribution (expected to be $1.15$ SNU
based upon the standard solar model~\cite{BP98}).

For all these reasons, we will consider in the remainder of this paper 
theories in which neutrino oscillations change the
shape of the neutrino energy spectra.

\section{INGREDIENTS AND TECHNIQUES}
\label{technique}

Many authors have reported  the 
results of refined 
studies  of the  MSW
\cite{hata95,fogli95,berezinsky95,HL,KPanal,Krauss,Kuo,fio,maris97} 
and the
vacuum \cite{KPanal,Krauss,fio,rossi} solutions of the solar 
neutrino problems. 
The techniques for this analysis are therefore well documented in the
literature and we only note briefly here those aspects of the
calculation that are often not treated in the optimal manner in the
literature or for which less accurate data are sometimes used.
Our $\chi^2$ analysis of the data follows closely the prescriptions in
\cite{brighter,howwell}. We adopt the procedures 
of Fogli and Lisi (see Ref.~\cite{fogli95}) in including theoretical
errors. The uncertainties in the input model parameters that influence
the neutrino fluxes are taken from Refs.~\cite{BP98,BP95}.
We use the improved 
neutrino interaction cross sections for each detector given
in Refs.~\cite{bahcallga97,bahcalletal96,bahcall88,sirlin}
and the  neutrino spectra given in Refs.~\cite{bahcallga97,balisi}.
We include the published  energy resolution and trigger efficiency  of
the Kamiokande detector \cite{kamiokande} 
and  SuperKamiokande\cite{superkamiokande300,superkamiokande374,superkamiokande504}.
For the MSW solution, we use the analytical
description of the neutrino survival probabilities from \cite{KP88}
which allows the averaging over the neutrino production regions and
the neutrino spectra to be done  accurately  with a
reasonable amount of computer time.

We obtain allowed regions in $\Delta m^2$ -
$\sin^22\theta$ parameter space by finding the minimum $\chi^2$ and
plotting contours of constant $\chi^2 = \chi^2_{min} + \Delta\chi^2$
where $\Delta\chi^2 = 5.99$ for 95\% C.L. and 9.21 for 99\%
C.L. .
In all the figures in this paper, we show results at the $99$\% C.L.

We describe the statistical analysis in detail in the Appendix.

\section{FITS TO THE AVERAGE EVENT RATES}
\label{averagerates}

In this section, we determine the allowed range of oscillation
solutions using only the total event rates in the ClAr (Homestake),
GALLEX, SAGE,  and SuperKamiokande experiments.\footnote{We have
carried out identical calculations including both the Kamiokande and the
SuperKamiokande rates. The results are essentially unchanged from what
we find including only the SuperKamiokande rate, since the quoted
uncertainty in the Kamiokande rate is much larger than the uncertainty
in the SuperKamiokande rate.} 

The average event rates in these four  solar neutrino
experiments, summarized in Table~\ref{datarates},
are robust and seem unlikely to change significantly.  The results of
the chlorine experiment have been summarized in detail
recently~\cite{homestake}; the measured value has been stable for 
two decades and
is known relatively precisely.  The Kamiokande~\cite{kamiokande} and 
SuperKamiokande~\cite{superkamiokande300,superkamiokande374,superkamiokande504} experiments are in agreement to within
$1\sigma$ and the precision of the SuperKamiokande experiment is now
very high.  We have used the more precise value of the SuperKamiokande
experiment to represent the total rate above $6.5$ MeV in the water
Cherenkov experiments. Both the GALLEX~\cite{GALLEX} and 
the SAGE~\cite{SAGE} experiments  have yielded accurate
measurements and the efficiency of both detectors has been tested with
 $^{51}$Cr sources.  We have used the weighted average of the measured
rates in the two gallium detectors.

We present in Sec.~\ref{ss:mswrates} the allowed range of solutions involving
MSW oscillation into active or sterile neutrinos.  In 
Sec.~\ref{ss:vacrates}, 
we present the corresponding results for vacuum oscillations.
Sec.~\ref{ss:arbitrary} describes the dependence of the inferred 
oscillation parameters on the most uncertain input parameter, the
low-energy cross section factor for the ${\rm ^7Be}(p,\gamma){\rm ^8B}$ reaction.
We show in Sec.~\ref{ss:constant} that energy-independent oscillations are
unacceptable at the $99.8$\% C.L. . Finally, we summarize in
Sec.~\ref{ss:ratessummary} 
our results on the analysis of the average event rates.

The calculation of the allowed range of predicted ${\rm ^7Be}$ flux (to be
measured in the BOREXINO experiment~\cite{borexino}), the
demonstration 
that a
constant suppression factor is a disfavored description, 
and the evaluation of the
dependence of the inferred neutrino parameters on the 
low-energy cross section factor for the ${\rm ^7Be}(p,\gamma){\rm ^8B}$ reaction
are special features of this section.

\subsection{MSW solutions}
\label{ss:mswrates}

\subsubsection{\it Active neutrinos}
\label{sss:mswactive}
The best fit is obtained for the small mixing angle (SMA) solution:

\begin{mathletters}
\label{SMAall}
\begin{eqnarray}
\label{SMApara}
\Delta m^2  &=&  5.4\times 10^{-6} {\rm eV}^2 ,\\ 
\sin^22\theta  &=&  6.0\times 10^{-3} ,
\label{SMAparb}
\end{eqnarray}
\end{mathletters}
which has $\chi^2_{\rm min} = 1.7$.  
There are two more local minima of $\chi^2$.  The best fit for the
well known large mixing angle (LMA) solution occurs at
\begin{mathletters}
\label{LMAall}
\begin{eqnarray}
\label{LMApara}
\Delta m^2  &=&  1.8\times 10^{-5} {\rm eV}^2 ,\\
\sin^22\theta  &=& 0.76 ,
\label{LMAparb}
\end{eqnarray}
\end{mathletters}
with $\chi^2_{\rm min} = 4.3$.  There is also a less probable
solution, which we refer to as the LOW solution (low probability, low
mass), at~\cite{krastev93,BaltzW3} 
\begin{mathletters}
\label{LOWall}
\begin{eqnarray}
\label{LOWpara}
\Delta m^2 & = &7.9\times 10^{-8} {\rm eV}^2 ,\\
\sin^22\theta & = & 0.96 .
\label{LOWparb}
\end{eqnarray}
\end{mathletters}
with $\chi^2_{\rm min} = 7.3$. 
The LOW solution is acceptable 
at the 99\% C.L.,  but is not acceptable at the 95\% C. L. .
To find an appreciable probability for the LOW solution, one
must include the regeneration effect~\cite{brighter}. 

How do the results given in Eq.~(\ref{SMAall})-Eq.~(\ref{LOWall})
differ from our pre-SuperKamiokande (1997) analysis? (The
1997 study, Ref.~\cite{brighter},  was carried out before either the
Bahcall-Pinsonneault 98 solar model   or   the
SuperKamiokande experimental 
results were available.)  There are no very large changes
in the best-fit values for either the mass differences or the mixing 
angles of the three solutions, although the $\chi^2$ fits are 
less good (but well within the statistical uncertainties,
cf. Eqs. (1)--(3) of Ref.~\cite{brighter}). However, the best-fit
neutrino parameters 
are shifted towards smaller mixing angles for the SMA and towards
larger mixing angles for the LMA and 
LOW solutions.  For the SMA
solution, this has the important result that the expected Day-Night
asymmetry is greatly decreased (see Sec.~\ref{zenith}.)
 
Figure~\ref{fig:mswrates} shows the 99\% C.L. allowed regions in  
the plane defined by $\Delta m^2$ and $\sin^2 2 \theta$.  
The black dots within
each allowed region indicate the position of the local best-fit point
in parameter space.
The results shown in Fig.~\ref{fig:mswrates}
were calculated using the predictions of
the 1998 standard solar model of Bahcall and
Pinsonneault\cite{BP98}, 
which includes helium and heavy element diffusion and uses the recent
reevaluation of solar fusion cross sections\cite{adelberger98};
the shape of the allowed contours depends only slightly upon the 
assumed solar model (see Fig.~1 of Ref.~\cite{howwell}).

The BOREXINO experiment will measure the $\nu-e$ scattering rate
 for the ${\rm 0.86~MeV~^7Be}$
line.  We have calculated the allowed range for the scattering rate 
at $99$\% C.L. for the different MSW solutions discussed above. The
rates are given in the following equation relative to the 1998 Bahcall and 
Pinsonneault standard model \cite{BP98}
are: 
\begin{mathletters}
\begin{equation}
\frac{\langle\phi\sigma\rangle_{\rm ^7Be~SMA}}{\langle\phi\sigma\rangle_{\rm
BP98}} = 0.23^{+0.24}_{-0.01} 
\label{eq:be7sma}
\end{equation}
\begin{equation}
\frac{\langle\phi\sigma\rangle_{\rm ^7Be~LMA}}{\langle\phi\sigma\rangle_{\rm
BP98}} = 0.59^{+0.15}_{-0.18} ~, 
\label{eq:be7lma}
\end{equation}
\begin{equation}
\frac{\langle\phi\sigma\rangle_{\rm ^7Be~LOW}}{\langle\phi\sigma\rangle_{\rm
BP98}} = 0.59^{+0.06}_{-0.08} ~.
\label{eq:be7low}
\end{equation}
\label{eq:Be7activeMSW}
\end{mathletters}

\subsubsection{\it Nuclear physics uncertainties}
\label{sss:mswnuclear}

After more than $35$ years of progressively more accurate measurements
of input parameters and more precise solar modeling, 
the largest recognized uncertainties that afflict the prediction of
solar neutrino fluxes are associated with the laboratory cross sections at low
energies for the crucial ${\rm ^3He}(\alpha,\gamma){\rm ^7Be}$ and the
${\rm ^7Be}(p,\gamma){\rm ^8B}$
reactions~\cite{BP98}.  The uncertainty in the ${\rm ^3He-^4He}$
cross section is $9.4$\%, $1\sigma$\cite{adelberger98}. 
The uncertainty in the ${\rm ^7Be} + p$
reaction is asymmetric and is, on average, $10.6$\%\cite{adelberger98}. 
The flux of  ${\rm ^8B}$ neutrinos is directly proportional to
the rate of the ${\rm ^7Be} + p$ reaction and the flux of the ${\rm ^7Be}$
neutrinos is approximately linearly proportional to the rate of the 
${\rm ^3He-^4He}$ reaction.

How much do the uncertainties in the nuclear physics parameters affect
the accuracy with which one can determine neutrino parameters?
In order to answer this question, we have
calculated the allowed regions for MSW solutions, analogous to those
considered in Sec.~\ref{sss:mswactive} , but we assumed in the present case 
either that the uncertainty in the 
${\rm ^3He}(\alpha,\gamma){\rm ^7Be}$ or  the ${\rm
^7Be}(p,\gamma){\rm ^8B}$ reaction was
equal to zero.  

Figure~\ref{fig:mswratesS0} shows that the MSW  allowed
regions are reduced only slightly when the
uncertainties associated with either the ${\rm ^3He}(\alpha,\gamma){\rm
^7Be}$ (cross-section factor $S_{34}$) or
the ${\rm ^7Be}(p,\gamma){\rm ^8B}$ reaction 
(cross-section factor $S_{17}$)
are artificially decreased to zero.
Comparing Fig.~\ref{fig:mswratesS0} and Fig.~\ref{fig:mswrates}, we
see that 
in both cases the allowed regions for MSW solutions remain comparable
in size to what they are calculated  to be with realistic estimates of
the nuclear physics uncertainties. The reason that the allowed regions
are not particularly sensitive to the uncertainties in any one nuclear
parameter is that there are a number of roughly comparable
uncertainties from different input parameters~\cite{BP98}.

We conclude that one cannot greatly increase the accuracy with which
MSW parameters can be determined by greatly decreasing the uncertainty
in any single input parameter.

\subsubsection{\it Sterile neutrinos}
\label{sss:mswsterile}

How are the results given above changed if the oscillations involve sterile
neutrinos?  

Figure~\ref{fig:mswratesst} shows that if the oscillations are
between an electron-type neutrino and a sterile neutrino then the LMA
and the LOW solutions are not allowed at the 99\% C.L. ; only the SMA
solution is possible . 
The LMA and LOW solutions are ruled out at a very high C.L.; the value of 
$\chi^2({\rm min}) = 19.0~(17.0) $ for the LMA (LOW) solutions. 
 However, the
SMA solution is still allowed, with $\chi^2({\rm min}) = 1.7$ (which
is the same as 
for the best-case with active neutrinos)  and the best-fit solution is 

\begin{mathletters}
\begin{equation}
\Delta m^2  =  4.3\times 10^{-6} \, {\rm eV}^2 ,
\label{eq:smaratesa}
\end{equation}
\begin{equation} 
\sin^22\theta  =  6.9\times 10^{-3} ~.
\label{eq:smaratesb}
\end{equation}
\label{eq:smaratesst}
\end{mathletters}
The mass difference for the solution involving 
sterile neutrinos, Eq.~(\ref{eq:smaratesa}), is slightly smaller than
the mass difference found for oscillations between active neutrinos
 (cf. Eq.~\ref{SMApara}), but the mixing angle for sterile neutrinos,
Eq.~\ref{eq:smaratesb}, is slightly larger than for active neutrinos
 (cf. Eq.~\ref{SMAparb}).

The suppression of the neutrino fluxes in the case of SMA conversion
into sterile neutrinos is somewhat similar to the arbitrary reductions
of the neutrino fluxes considered in Sec.~\ref{ss:mit}.  The flux
of $pp$ neutrinos is unsuppressed, the ${\rm ^7Be}$ neutrino flux as well as
other fluxes of intermediate energies are strongly suppressed, and the
${\rm ^8B}$ neutrino flux is moderately suppressed. The fit to the solar
neutrino experiments using oscillations into  sterile
neutrinos is much better than the fit to the experiments made using
an arbitrary linear combination of undistorted solar neutrino fluxes
 (cf. Sec.~\ref{ss:mit}).  The main difference between the
suppression due to sterile neutrinos and the suppressions considered in  
Sec.~\ref{ss:mit} is that oscillations into sterile neutrinos
suppress the lower energy part of the ${\rm ^8B}$ neutrino energy
spectrum.  This result is an illustration of our claim made in
Sec.~\ref{ss:discussionsm} that the evidence from the total
experimental rates 
suggests the existence of a distortion of the solar neutrino energy spectrum.

What should one expect for BOREXINO if oscillations involve sterile 
neutrinos? The allowed range for the ${\rm ^7Be}$ electron scattering rate
is 
\begin{equation}
\frac{\langle\phi\sigma\rangle_{\rm ^7Be~sterile}}{\langle\phi\sigma\rangle_{\rm
BP98}} = 0.009^{+0.244}_{-0.005}  .
\label{eq:be7sterilesma}
\end{equation}
If oscillations occur to sterile neutrinos, the rate observed in
BOREXINO will be---for almost the entire range of allowed
parameters---less 
than the lowest expected rate if oscillations occur to active
neutrinos (cf. Eq.~\ref{eq:smaratesst} ).

If the Caltech normalization for the ${\rm ^8B}$ production cross section 
($S_{17} = 22.4 ~{\rm kev~b}$) is used instead of the INT
normalization, the allowed regions are shifted only slightly from what
they are in Figure~\ref{fig:mswrates} and 
Figure~\ref{fig:mswratesst}.  The best estimate solutions
for $\delta m^2$ are changed by less than $7$\% relative to the values
given in Eqs.~(\ref{SMAall})--(\ref{LOWall}) and 
Eq.~(\ref{eq:smaratesst}).

\subsection{Vacuum neutrino oscillations}
\label{ss:vacrates}

Figure~\ref{fig:vacrates} shows for vacuum neutrino oscillations 
the broad region of solutions allowed at 99\% C.L. .
The best-fit vacuum solution is

\begin{mathletters}
\begin{equation}
\Delta m^2  =  8.0\times 10^{-11} {\rm eV}^2 ,
\label{eq:vacratea}
\end{equation}
\begin{equation} 
\sin^22\theta  = 0.75 ~,
\label{eq:vacrateb}
\end{equation}
\label{eq:vacrates}
\end{mathletters}
which has a $\chi^2_{\rm min} = 4.3$.

What will BOREXINO observe if vacuum neutrino oscillations occur?  The
allowed range for vacuum oscillations of the ${\rm ^7Be}$ line 
is very large if the only constraints imposed are consistent with the
total observed rates. We find at $99$\%
C.L. that 
\begin{equation}
\frac{\langle\phi\sigma\rangle_{\rm ^7Be~vacuum}}{\langle\phi\sigma\rangle_{\rm
BP98}} = 0.46^{+0.46}_{-0.18}  ~.
\label{eq:be7vac}
\end{equation}

There is no allowed solution at the $99.7$\% C.L. for 
vacuum neutrino oscillations
between $\nu_e$ and a sterile neutrino.  The solution involving sterile
neutrinos has $\chi^2({\rm min}) = 12.0$.

\subsection{Arbitrary ${\rm ^8B}$ neutrino flux}
\label{ss:arbitrary}

The value of the 
${\rm ^8B}$ neutrino flux calculated 
in the standard solar model (cf. Ref.~\cite{BP98}) is more uncertain
($+19$\% and $-14$\%, $1\sigma$) than
any of the other experimentally-important solar neutrino fluxes.
It is therefore useful to consider what constraints are placed upon
neutrino physics if the ${\rm ^8B}$ flux is treated as a free parameter
~\cite{krastevsmirnov94,hata95}.

\subsubsection{\it MSW solutions}
\label{sss:mswarbitrary}

Figure~\ref{fig:mswvarb8} shows the allowed parameter space for MSW
oscillations when the ${\rm ^8B}$ flux is allowed to take on arbitrary values.
The best-fit SMA solution is

\begin{mathletters}
\label{varSMAall}
\begin{eqnarray}
\label{varSMApara}
\Delta m^2  &=&  5.0\times 10^{-6} {\rm eV}^2 ,\\ 
\sin^22\theta  &=&  3.5\times 10^{-3} ,
\label{varSMAparb}
\end{eqnarray}
\end{mathletters}
which has $\chi^2_{\rm min} = 0.86$.
The best-fit for the LMA solution is
\begin{mathletters}
\label{varLMAall}
\begin{eqnarray}
\label{varLMApara}
\Delta m^2  &=&  1.6\times 10^{-5} {\rm eV}^2 ,\\
\sin^22\theta  &=& 0.57 ,
\label{varLMAparb}
\end{eqnarray}
\end{mathletters}
which has  $\chi^2_{\rm min} = 0.91$.  The LOW solution,
\begin{mathletters}
\label{varLOWall}
\begin{eqnarray}
\label{varLOWpara}
\Delta m^2 & = &7.9\times 10^{-8} {\rm eV}^2 ,\\
\sin^22\theta & = & 0.95 ,
\label{varLOWparb}
\end{eqnarray}
\end{mathletters}
has a much larger  $\chi^2_{\rm min} = 7.2$. 
For all three classes of MSW solutions the 
squared mass differences are
changed by less than $20$\%  if the ${\rm ^8B}$ neutrino flux is
treated as a free parameter (cf. Eq.~\ref{SMAall}--
Eq.~\ref{LOWall}),  although the values of $\sin^22\theta$
are changed by much large factors (a factor of two for the SMA solution).

Do the solar neutrino observations place useful limits on the value of
the cross section factor, $S_{17}$, for the production of ${\rm ^8B}$?

Figure~\ref{fig:minchivarb8} shows the minimum value of $\chi^2$
obtained using as constraints the rates of the solar neutrino
experiments but allowing the value of the ${\rm ^8B}$ flux to take on
arbitrary values. The allowed range at $99$\% C.L. is: 
$0.4 < S_{17}/S({\rm INT})_{17} < 2.0$, which is much broader than the
range allowed\cite{adelberger98} by direct laboratory experiments.

If oscillations between sterile neutrinos are considered, the
constraints set on $S_{17}$ by varying the solar  $^8$B 
flux and fitting  to the solar neutrino
data are again  much larger than the uncertainty in the laboratory
measurement of $S_{17}$ (cf. Ref~\cite{krastevpetcov96}).

\subsubsection{\it Vacuum oscillations}
\label{sss:vacarbitrary}

How are the inferred parameters for vacuum neutrino oscillations
affected by permitting arbitrary values for the assumed ${\rm ^8B}$ neutrino
flux, cf. Ref.~\cite{krastevpetcov96}?  

Figure~\ref{fig:vacvarb8} shows the expanded solution space
that is allowed for vacuum oscillations 
if the ${\rm ^8B}$ neutrino flux is unconstrained. Instead
of the rather limited parameter space that is permitted with the
standard solar model  value of the ${\rm ^8B}$ flux (see Fig~\ref{fig:vacrates}), 
the squared neutrino mass difference can span the entire 
two order of magnitude range from 
$\Delta m^2 = 4\times10^{-12} {\rm eV^2}$ to 
$\Delta m^2 = 5\times10^{-10} {\rm eV^2}$ if 
$S_{17}$ is allowed to vary arbitrarily.
The allowed range of the ${\rm ^8B}$ cross section factor at $99$\% C.L. is,
for vacuum oscillations, 
$0.4 < S_{17}/S({\rm INT})_{17} < 2.0$.

For completeness, we note that the best-fit values for the vacuum
solution and arbitrary ${\rm ^8B}$ flux are 
\begin{mathletters}
\begin{equation}
\Delta m^2  =  8.4\times 10^{-11} {\rm eV}^2 ,
\label{eq:vacvar8ba}
\end{equation}
\begin{equation} 
\sin^22\theta  = 0.98 ~,
\label{eq:vacvar8bb}
\end{equation}
\label{eq:vacvarrates}
\end{mathletters}
which has a shallow $\chi^2_{\rm min} = 0.94$.  The minimum occurs for 
a ${\rm ^8B}$ flux that is $1.9$ times the value of the predicted BP98
${\rm ^8B}$ flux.

This allowed range of $S_{17}$ is very similar for both the MSW and
the vacuum solutions.
In both cases, the  measurements by 
 SuperKamiokande determine the upper and 
lower limits on the allowed ${\rm ^8B}$ flux.  The lower limit corresponds
to very little conversion from $\nu_e$
to other neutrino types.  The lowest allowed value is
obtained by  reducing the best-estimate SuperKamiokande flux by 
the $3\sigma$ experimental uncertainty. Since the experimental
uncertainty is small ($1\sigma_{\rm exp} = 3$\%), the lower limit is close
to the SuperKamiokande best-estimate value. The upper limit is much
larger because essentially all of the ${\rm ^8B}$ flux
observed by SuperKamiokande is $\nu_\mu$ or $\nu_\tau$.  Since the
cross section~\cite{sirlin} for electron-neutrino scattering by $\nu_\mu$ or
$\nu_\tau$ is about six times smaller than the scattering cross
section for $\nu_e$, and $\nu_\mu$ and  $\nu_\tau$ are not detected in
the chlorine and gallium experiments, 
the upper limit for the total flux can be much larger (a factor of
five in practice) than the observed flux. 

\subsection{Energy-independent suppression of neutrino fluxes}
\label{ss:constant}

Many authors have considered particle-physics models which predict 
energy-independent reductions of the
solar neutrino fluxes (see, e. g.,
Ref.~\cite{nussinov76,harrison,foot,comforto,acker}} 
and 
references cited
therein).  In these scenarios, all neutrino fluxes are reduced by some
particle-physics mechanism by exactly the same factor. One can test
the goodness-of-fit of such scenarios by calculating the minimum
$\chi^2$ for this case~\cite{krastevpetcov97}.
Using the predictions and the uncertainties in
the BP98 solar model and the observed rates and their uncertainties
given in Table~\ref{datarates}, we have calculated the minimum $\chi^2$ for
different neutrino oscillation scenarios.

For arbitrary reduction factors $\alpha$, the best-fit value is 
$\alpha = 0.48$ for oscillation into active neutrinos and $\alpha =
0.50$ for sterile neutrinos. The minimum $\chi^2_{\rm min} = 12.0$ for active
neutrinos and $\chi^2_{\rm min} = 19.3$ for sterile neutrinos.  Since
there are two degrees of freedom in these cases, 
energy-independent oscillation into active
neutrinos is ruled out at the $99.8$\% C.L. and energy-independent
oscillation into sterile neutrinos is ruled out at the $99.99$\%C.L. .
These results are changed only slightly (strengthened slightly) if the
$^8$B flux is allowed to vary independently of all other fluxes. In
this case, we obtain for active neutrinos (sterile neutrinos) 
$\chi^2_{\rm min} = 11.0$ ($\chi^2_{\rm min} = 19.3$) for $1$ d.o.f.,
which is excluded at the $99.8$\% C.L. ($99.999$\% C.L.) .

For the model considered in Ref.~\cite{harrison}, $\alpha = 5/9$.  In this
case, $\chi^2_{\rm min} = 14.3$  (3 d.o.f.), and the model is ruled
out at the $99.8$\% C. L. .

\subsection{Summary of analysis of rates}
\label{ss:ratessummary}

The principal results of this section are displayed in
Fig.~\ref{fig:mswrates}--Fig.~\ref{fig:vacvarb8}, 
except for the BOREXINO predictions which are given
in Eq.~(\ref{eq:Be7activeMSW}), Eq.~(\ref{eq:be7sterilesma}), and 
Eq.~(\ref{eq:be7vac}), and the demonstration in Sec.~\ref{ss:constant} that 
constant-reduction solutions are disfavored. 
For the reader's convenience, we also present the numerical values of
the best-fit neutrino mixing angles and mass differences for each case
considered.

In all of the calculations
shown in the figures, we have used the INT normalization for the ${\rm
^8B}$
production cross section, $S_{17}$. However, this cross section is
relatively poorly known and, unfortunately, 
at present it is largely a matter of judgment as to
which normalization, INT or CIT,  
is most appropriate.  We have therefore presented results for both
normalizations; the difference between neutrino parameters determined
with the INT and the CIT normalizations is  an indication 
of the magnitude of the uncertainties in neutrino parameters that are
caused by uncertainties in the basic nuclear physics data.

Particle physics solutions in which the suppression of the electron
neutrino flux is independent of the neutrino energy are ruled out at
more than the $99$\% C. L. for oscillations into both active and
sterile neutrinos (see Sec.~\ref{ss:constant}).

Table~\ref{tab:minchi} presents the minimum $\chi^2$ for the different
neutrino oscillation scenarios that we have considered in this
section. The primary
entries in the table represent $\chi^2({\rm min})$ with the INT
normalization for $S_{17}$; the entries enclosed in parentheses
were computed using the CIT normalization for $S_{17}$. 
The INT and CIT normalizations give about equally good fits to the
observed rates in solar neutrino experiments, with the INT
normalization slightly favoring the SMA MSW solution and the CIT
normalization slightly favoring the LMA solution (see also
Fig.~\ref{fig:minchivarb8}). 
Indeed, 
the larger ${\rm ^8B}$ flux, the CIT
normalization,
implies a stronger suppression by oscillations at higher energies, 
and consequently, a flatter
dependence of the suppression factor on energy than is implied by the 
INT normalization.

\section{ZENITH-ANGLE DEPENDENCE OF RATES}
\label{zenith}

If MSW oscillations occur, the observed event rates in solar
neutrino experiments can depend upon which region, if any,  of the
earth that the
neutrinos traverse before reaching the detector. At times when the
sun is below the horizon, $\nu_\mu$ or $\nu_\tau$ coming from the sun
can be re-converted in the earth into the more easily detected
$\nu_e$. Thus for a certain range of neutrino parameters the observed
event rate will depend upon the zenith angle  of the sun,
i.e., the angle that the sun makes with respect to the direction of
the zenith at the position of the neutrino detector. This is known as
the earth regeneration effect \cite{earthreg,BaltzW,BaltzW2}.

\subsection{Expectations for survival probabilities}
\label{ss:expectations}

Figure~\ref{fig:survival} shows the computed survival probabilities
for 
electron type neutrinos as a function of energy for the
day (no regeneration), the night (with regeneration), and the annual
average. 
The survival probabilities computed here for the best-fit points 
differ relatively  little from the values calculated earlier
 (see Fig.~4 of Ref.~\cite{brighter}),
despite the fact that in the present study we have 
taken account of  the measured
SuperKamiokande rate, the somewhat different 
fluxes from the BP98 model, the improved
estimates for  nuclear fusion cross sections, and other updated data.
The changes in the best-fit oscillation parameters caused by the
use of the new data are relatively small (see
Sec.~\ref{averagerates}).
The principal difference is that with the present parameters the
expected Day-Night difference for the SMA solution 
is extremely small, not even visible in 
Fig.~\ref{fig:survival}.  Note, however,  that a
potentially detectable Day-Night difference is predicted for the LMA
solution at the higher neutrino energies. The Day-Night difference is
expected to be relatively large for the LOW solution, but only at
energies below $1$ MeV. The Day-Night difference predicted by 
 the LOW solution would be detectable by the BOREXINO experiment,
which will observe the $0.86$ MeV neutrino line from ${\rm ^7Be}$ electron capture.

\subsection{SuperKamiokande result}
\label{ss:superkresultdn}

The SuperKamiokande Collaboration has given a preliminary value for
the difference between the event rates at night and during the day.
After only 504 days of data, they obtain an initial estimate for this
Day-Night asymmetry, $A$,  of~\cite{superkamiokande300,superkamiokande504}:

\begin{equation}
A = { {D - N} \over {D + N} } = -0.023 \pm 0.020 ({\rm stat}) \pm 0.014
({\rm syst}) .
\label{eq:daynight}
\end{equation}
This estimate applies for events in which the recoil electron has an
energy of at least $6.5$ MeV.  
The difference shown in Eq.~(\ref{eq:daynight}) is in the direction
that would be expected from regeneration in the earth (the sun is apparently
brighter at night in neutrinos), but is small and is 
not statistically significant.

The SuperKamiokande 
Collaboration has also
given~\cite{superkamiokande300,superkamiokande504}
 a preliminary 
distribution of the
event rates versus zenith angle in which the rates and their errors
are plotted in 10 angular bins.
 
\subsection{Constraints on neutrino parameters}
\label{ss:constaintsdn}

We have determined the constraints placed upon neutrino mixing
parameters using  the preliminary Day-Night asymmetry
(Eq.~\ref{eq:daynight}) and also the 10-bin zenith angle distribution.
We use the techniques and results contained in our previous detailed
discussion of  the expected zenith-angle dependence of solar event
rates~\cite{brighter} to analyze the 10 bin measurements
of the zenith angle distribution reported by the SuperKamiokande
Collaboration for the first 504 days of 
observations~\cite{superkamiokande504}.

Figure~\ref{fig:mswangular}a shows the results of a combined MSW
$\chi^2$ fit with active neutrinos 
of the SuperKamiokande zenith angle dependence and the
total rates of the Homestake, GALLEX, SAGE, and SuperKamiokande experiments.
The region in the figure that is excluded by the zenith angle
dependence is shaded lightly  and, almost touches the
darkly shaded LMA and SMA allowed regions for the combined zenith angle and
rate constraints.  Figure~\ref{fig:mswangular}b shows the
combined solutions using the SuperKamiokande Day-Night asymmetry and the
four total rates. 
As shown previously from simulated data~\cite{brighter}, the Day-Night
asymmetry is more sensitive to the LMA solution and the
zenith-angle distribution is more sensitive to the SMA solution.
We see from Fig.~\ref{fig:mswangular}b that the observations by 
SuperKamiokande of the Day-Night asymmetry have
eliminated a sizable fraction, almost a  half, 
of the parameter space for the LMA
solution that is allowed if one only considers the total 
rates (cf. Fig.~\ref{fig:mswrates}). 
A  small part 
(with $\sin^2 2\theta > 10^{-2}$) of the parameter space of the SMA
solutions is excluded  by the  measured zenith-angle dependence.

Figure~\ref{fig:mswangularsterile}a and Figure~\ref{fig:mswangularsterile}b
show the analogous regions 
for MSW oscillations into sterile
neutrinos. The only MSW solution for oscillation into sterile
neutrinos is the SMA. The regeneration
effect is not as effective a test of the sterile neutrino oscillation 
solution as it is for
oscillation into active
neutrinos (cf. Fig.~\ref{fig:mswangularsterile} and Fig.~\ref{fig:mswrates}).

\section{PRELIMINARY SPECTRAL SHAPE}
\label{spectralshape}

If minimal standard electroweak theory is correct, 
the shape of the ${\rm ^8B}$ neutrino energy spectrum is independent of all solar
influences to an accuracy of $1$ part in $10^5$~\cite{bahcall91}.
The shape of the neutrino spectrum determines the shape of the
 recoil electron energy spectrum produced by neutrino-electron
scattering in the detector.
Therefore, any  departure of the observed electron recoil
energy spectrum from the shape predicted by standard electroweak
theory would be a ``smoking gun'' indication of new physics.
In this section, we compare the preliminary recoil electron energy
spectrum reported by SuperKamiokande~\cite{superkamiokande504} with the
results calculated using the standard (undistorted)
${\rm ^8B}$~\cite{balisi} neutrino spectrum.  We also compare the observed
energy spectrum with recoil energy spectra calculated assuming MSW or
vacuum neutrino oscillations and use these results to constrain the
allowed neutrino parameter space. 

\subsection{SuperKamiokande energy spectrum}
\label{ss:spectrum}

The SuperKamiokande Collaboration has made
available~\cite{superkamiokande504} preliminary data,
including estimated statistical and systematic uncertainties,
from 504 days of operation in which the recoil energy spectrum is 
divided into $16$ bins, with 15 bins having a 
width of $0.5$ MeV starting at $6.5$
MeV and continuing to $14$ MeV.
The final bin includes events with energies
from $14$ MeV to $20$ MeV and contains more counts than would be
expected from an undistorted $^8$B spectrum normalized at lower
energies.

\subsection{Fit to undistorted energy spectrum}
\label{ss:undistorted}

The minimum $\chi^2$ for the fit of the undistorted  energy spectrum to
the measured energy spectrum is $31$ for 15 D.O.F. . The fit with the 
undistorted spectrum is acceptable only at slightly less than 
the $1$\% C.L., consistent
with the results reported by the SuperKamiokande Collaboration at
Neutrino 98~\cite{superkamiokande504}.

\subsection{Fits to MSW and vacuum oscillations}
\label{ss:distorted}

The simplest test~\cite{refstraightline}
for a deviation from the standard recoil energy
spectrum is to investigate whether the ratio, $R$, of the observed to 
the standard  energy spectrum is a constant, which would be expected in
the absence of a distortion.  
We have therefore fit the ratio R to a
linear function using the measured~\cite{superkamiokande504} 
 number of events and their quoted
uncertainties in the 
16 bins (from $6.5$ MeV to $20$ MeV) and the standard electron recoil energy
spectrum~\cite{balisi} modified by the energy resolution of the
SuperKamiokande detector.  Thus

\begin{equation}
R ~=~ R_0 ~+~S_0*(W_e - 10\,{\rm MeV}),
\label{eq:ratioR}
\end{equation}
where $R_0$ represents the average event rate 
and $S_0$ is the average slope that measures the deviation of the
recoil electron energy spectrum from the undistorted shape. 
The total electron energy is  $W_e$.
The energy spectrum of recoil electrons is determined  by convolving 
 the neutrino spectrum ~\cite{balisi} with the calculated 
survival probability, the 
 neutrino-electron scattering 
  cross-sections~\cite{sirlin}, and the energy resolution 
function~\cite{superkamiokande504}.  
These integrations smooth the effect of distortions;
therefore, the expected distortion can be well 
described by a simple linear dependence.

Figure~\ref{fig:linefit} shows the result of this
calculation. 
The $1\sigma$, $2\sigma$, and 
$3\sigma$ allowed regions are  shown in the figure.
We have taken account of the bin correlations between the systematic
errors in the fashion explained in the Appendix.  
The best-fit point lies at $R_0 = 0.474$ and $S_0 = 0.0153$,
with $\chi^2_{\rm min} = 23.5$ for $14$ d.o.f..  Therefore, a straight
line is not a particularly good  fit to the data; it is acceptable at
the $5.3$\% C.L. .  The main reason that the fit is acceptable only at
a modest C.L. is that the last three bins indicate a deviation from a 
smooth extrapolation at lower energies.  This deviation could be real
or it could be a statistical fluctuation.

Figure~\ref{fig:linefit} also shows the ranges of the
predicted slopes ($S_0$) for different successful descriptions of the
solar neutrino total experimental 
rates, i.e., the SMA active and sterile, LMA, LOW, and vacuum oscillation
solutions described in Sec.~\ref{averagerates}.
All of the oscillation 
solutions shown in Fig.~\ref{fig:linefit} permit
a wide range of $R_0$ that includes the range measured by the
SuperKamiokande Collaboration.
For visual convenience, we have not included the horizontal error
bars, i. e., the uncertainties in   $R_0$, for the neutrino
oscillation solutions
shown  in Fig.~\ref{fig:linefit}.\footnote{The allowed range,
cf. Sec.~\ref{averagerates},  of 
$R_0$ is $0.38$ to $0.86$ for the SMA active solution, 
$0.33$ to $0.84$ for the SMA sterile neutrin solution,
$0.31$ to $0.55$ for
the LMA solution, and $0.42$ to $0.53$ for the LOW solution.}
Therefore, there are acceptable oscillation solutions of all four types that
describe both the total rates in the four solar neutrino experiments
and the recoil electron energy spectrum measured by SuperKamiokande.

How powerful a constraint is the spectrum shape in determining the
allowed neutrino oscillation parameters?

Figure~\ref{fig:spectrumMSWa}
shows, for both active and sterile neutrinos, 
the parameter region that is allowed  by considering only  the spectral
information.  Only a small fraction  of MSW parameter space is
consistent with the spectral data. 
It is instructive to compare the
regions allowed by the rates only, shown in  
Fig.~\ref{fig:mswrates} and Fig.~\ref{fig:mswratesst}, with the
regions shown in Fig.~\ref{fig:spectrumMSWa}, which are
 allowed by the spectrum data. 
The best-fit solutions considering only the spectrum, marked by dark points in
Fig.~\ref{fig:spectrumMSWa}, do not lie within the allowed regions for
the global solutions discussed in Sec.~\ref{global}.

Figure~\ref{fig:spectrumV} shows the region that is allowed for vacuum
oscillations by the information from the SuperKamiokande energy spectrum.
The dark circle shows the best-fit point.  Just as for MSW oscillations,
only a small fraction of the parameter space is allowed for vacuum
oscillations by the spectrum constraint. 
The complementarity of the  analysis that uses only the rates 
and the analysis that uses 
only the spectral data can be seen by comparing 
Fig.~\ref{fig:vacrates} and Fig.~\ref{fig:spectrumV}.

\section{GLOBAL SOLUTION: RATES, ZENITH-ANGLE DISTRIBUTION, 
AND ENERGY SPECTRUM}
\label{global}

We discuss in this section simultaneous fits to all the available
data. We consider solutions that describe the total 
rates in the four experiments (chlorine, SuperKamiokande, GALLEX, and
SAGE) plus either the energy spectrum measured by SuperKamiokande or
both the energy spectrum and the Day-Night asymmetry (or 
zenith-angle dependence) measured by
SuperKamiokande. 

We begin by showing in Sec.~\ref{ss:arbglobal}
that an arbitrary combination of
undistorted solar neutrino energy spectra is, independent of any
astrophysical considerations,  ruled out at the
$3.5\sigma$ level. We next show in Sec.~\ref{ss:mswglobal}
that the best global fit for the MSW
solutions is acceptable at the $7$\% C.L. for active neutrinos ($8$\%
C.L. for sterile neutrinos). The best-fit global solution 
 is very close the best-fit
SMA solution when only the total experimental rates are considered. 
Finally, we show in Sec.~\ref{ss:vacglobal} that the best global fit
for vacuum oscillations is acceptable at the $5$\% C.L. .

The results of the simultaneous fits 
to all of the available data of the neutrino predictions are shown in
Fig.~\ref{fig:global}--Fig.~\ref{fig:globalV}, which present the
allowed regions for the different oscillation scenarios. 

\subsection{Global fits: arbitrary undistorted fluxes}
\label{ss:arbglobal}

What is the best fit to the total rates plus the SuperKamiokande
spectrum and Day-Night asymmetry if we allow  
arbitrary values, subject only to the luminosity constraint~\cite{howwell}, 
 for the $pp$, ${\rm ^7Be}$, ${\rm ^8B}$, and CNO   
fluxes ?  The minimum $\chi^2$ 
satisfies

\begin{equation}
\chi^2_{\rm minimum} \hbox{(3 rates +  spectrum + D/N;  
~arbitrary $pp$, ${\rm ^7Be}$, ${\rm ^8B}$, ${\rm ^{13}N}$, ${\rm ^{15}O}$)
 = 39.2\ , }
\label{eqarbglobal}
\end{equation}
for $15$ d.o.f.\footnote{There are 3 d.o.f associated with the
rates, 15 d.o.f associated with the spectra shape which has one
overall normalization parameter that is variable, and 1 d.o.f.  for the
Day-Night asymmetry. All 5 fluxes are allowed to vary freely, subject
only to the luminosity constraint.} 
This result is excluded at the $99.94$\%
C.L., which corresponds to a $3.5\sigma$ deviation from minimal
electroweak theory.

\subsection{Global fits: MSW solutions}
\label{ss:mswglobal}

Fig.~\ref{fig:global}a displays the results of imposing on the MSW solutions
the combined constraints of the total rates
and the SuperKamiokande spectrum.  
The best-fit MSW solution considering both total rates in the four
experiments and the
electron recoil energy spectrum measured by SuperKamiokande has 
\begin{mathletters}
\begin{equation}
\Delta m^2  =  5.4\times 10^{-6} {\rm eV}^2 ,
\label{eq:mswratesspeca}
\end{equation}
\begin{equation} 
\sin^22\theta  = 6.3\times 10^{-3}~.
\label{eq:mswratespecb}
\end{equation}
\label{eq:mswratespec}
\end{mathletters}
The minimum $\chi^2_{\rm min} = 26.5$, which is acceptable at the
$7$\% C.L. (for $17$ d.o.f.), not a very good fit.
The best-fit solution is very close to the SMA solution when only rates are
considered (cf. Eq.~\ref{SMAall}).  Including the spectrum in addition to
the total rates, eliminates (at $99$\% C.L.) the entire LOW solution
and a large portion of the LMA solution at higher $\Delta m^2$. 
The inclusion of the spectral constraint also eliminates for the SMA
solution part of the region at smaller mixing angles that is allowed
if only the total rates are considered.

Fig.~\ref{fig:global}b shows the allowed region for MSW parameters
when the constraints from the SuperKamiokande 
zenith-angle distribution is included
together with the constraints from the four measured total rates and
the SuperKamiokande electron recoil energy 
spectrum.\footnote{We have not shown the combined fit including the 
Day-Night asymmetry, the spectrum constraint, and the total
experimental rates since the excluded region is very similar to what 
appears in Fig.~\ref{fig:global}b.  We chose to display  the fit
with the zenith-angle distribution since this provides the best
restriction  on the SMA solution.}  The best-fit
solution is almost identical to what is obtained for the rates only
case and for the case of rates plus zenith-angle constraint, namely, 

\begin{mathletters}
\begin{equation}
\Delta m^2  =  5.4\times 10^{-6} {\rm eV}^2 ,
\label{eq:mswratesspeczenitha}
\end{equation}
\begin{equation} 
\sin^22\theta  = 5.5\times 10^{-3}~.
\label{eq:mswratespeczenithb}
\end{equation}
\label{eq:mswratespeczenith}
\end{mathletters}
The minimum $\chi^2_{\rm min} = 37.2$, which again is acceptable at the
$7$\% C.L. (for $26$ d.o.f.).
Only the SMA solution survives at the $99$\% C.L. when the
zenith-angle and the spectrum constraints are added to the constraints
of the total experimental rates. The LMA and the LOW solution are
marginally ruled out ($\chi^2_{\rm min} = 47 (49)$ for the LMA (LOW)
solution).

The Day-Night asymmetry parameter, $A$, which is 
 defined by Eq.~(\ref{eq:daynight}), lies in the range

\begin{equation}
  -0.0048 < A < 0.025 ~,
\label{eq:globaldaynighta}
\end{equation}
for active neutrinos when all three sets of constraints are applied.
The best-fit solution predicts a Day-Night asymmetry whose absolute
value is less than $0.1$\%. 

How is the global solution changed if oscillations involve sterile
neutrinos? Figure~\ref{fig:globalMSWst} shows the allowed solution
space when the measured total rates, energy spectrum, and zenith-angle
dependence are all included as constraints. The best-fit solution is
 very close to what is obtained from the rates only (or the rates
plus spectrum constraints) for sterile neutrinos. We find

\begin{mathletters}
\begin{equation}
\Delta m^2  =  4.0\times 10^{-6} \, {\rm eV}^2 ,
\label{eq:globalsmasta}
\end{equation}
\begin{equation} 
\sin^22\theta  =  6.9\times 10^{-3} ~.
\label{eq:globalsmastb}
\end{equation}
\label{eq:globalsmast}
\end{mathletters}
The minimum $\chi^2_{\rm min} = 36.5$, which  is acceptable at the
$8$\% C.L. (for $26$ d.o.f.), slightly better than for active neutrinos.
For sterile neutrinos, the Day-Night asymmetry parameter, $A$, 
lies in the range

\begin{equation} 
-0.0031 < A_{\rm sterile} < 0.007 ~.
\label{eq:globaldaynightst}
\end{equation}
The best-fit solution predicts a value for the asymmetry parameter
whose absolute value is less than $0.1$\%.

\subsection{Global fits: vacuum oscillations}
\label{ss:vacglobal}

Fig.~\ref{fig:globalV} shows the allowed regions for the vacuum
oscillation  parameters when the constraints from the rates,   the
spectrum shape, and the Day-Night asymmetry are all  included.  
The best-fit vacuum oscillation solution
has 
\begin{mathletters}
\begin{equation}
\Delta m^2  =  6.5\times 10^{-11} {\rm eV}^2 ,
\label{eq:vacratesspeca}
\end{equation}
\begin{equation} 
\sin^22\theta  = 0.75~.
\label{eq:vacratesspecb}
\end{equation}
\label{eq:vacratesspec}
\end{mathletters}
The minimum $\chi^2_{\rm min} = 28.4$, which  is acceptable at the
$6$\% C.L. (for $18$ d.o.f.). This value of $\chi^2_{\rm min}$
 is only slightly below the value
of $\chi^2 = 30.5$ that is found for the best-fit point in the
`spectrum-only' analysis (cf. Fig.~\ref{fig:spectrumV}).
The inclusion of the spectral constraints reduces considerably the
domain of 
allowed solutions for vacuum oscillations (cf. Fig.~\ref{fig:globalV}
and Fig.~\ref{fig:vacrates}).

\subsection{Global fits: predicted ${\rm\bf ^7Be}$ $\nu-e$
scattering rates}
\label{ss:be7}

We have calculated the range of allowed ${\rm ^7Be}$ neutrino-electron
scattering rates  ($0.86$
MeV line) that is consistent with all of the available solar neutrino
data: total event rates, Day-Night asymmetry, and spectrum shape. The
results are very similar to what was obtained earlier in the
discussion of the rates only, see  
Eq.~(\ref{eq:Be7activeMSW}), Eq.~(\ref{eq:be7sterilesma}), and 
Eq.~(\ref{eq:be7vac}).  
We conclude that the neutrino oscillation predictions for the
BOREXINO experiment are robust.  We find, in particular, that the
global range predicted by the MSW solution with active neutrinos is

\begin{equation}
\frac{\langle\phi\sigma\rangle_{\rm ^7Be~SMA}}{\langle\phi\sigma\rangle_{\rm
BP98}} = 0.23^{+0.24}_{-0.01} .
\label{eq:globalbe7sma}
\end{equation}
The corresponding range for MSW oscillations into sterile neutrinos is
\begin{equation}
\frac{\langle\phi\sigma\rangle_{\rm ^7Be~sterile}}{\langle\phi\sigma\rangle_{\rm
BP98}} = 0.006^{+0.25}_{-0.002}  .
\label{eq:globalbe7sterilesma}
\end{equation}
For vacuum oscillations, 
\begin{equation}
\frac{\langle\phi\sigma\rangle_{\rm ^7Be~vacuum}}{\langle\phi\sigma\rangle_{\rm
BP98}} = 0.45^{+0.33}_{-0.11}  ~.
\label{eq:globalbe7vac}
\end{equation}

The principal change caused by the imposition of the additional
constraints due to the measured spectrum shape and Day-Night asymmetry
is a modest shrinking of the allowed range for the vacuum solutions.

\subsection{Global fits: predicted  ${\rm\bf ^7Be}$ $\nu_e$ fluxes}
\label{ss:nuebe7}

Raghavan~\cite{raghavan97} has recently proposed a flavor-specific
(neutrino-absorption) experiment that measures the total flux of
$0.86$ MeV 
$\nu_e$ reaching the earth from $^7$Be electron captures in the sun. 
The preliminary name for this experiment is LENS.

We have calculated the range of allowed $^7$B $\nu_e$ survival
probabilities for comparison with the results of a future LENS
experiment. We again considered neutrino parameters consistent with all the
available solar neutrino data: total event rates, Day-Night asymmetry,
and spectrum shape. We give below the predicted values  for 
the survival probability, $P$, for an
electron-type $0.86$ MeV neutrino created in sun to remain an
electron-type neutrino when it reaches the terrestrial target.
For MSW oscillations into active neutrinos we find 

\begin{equation}
{P_{\rm ^7Be~SMA}} = 0.02^{+0.30}_{-0.01} .
\label{eq:globalebe7sma}
\end{equation}
The corresponding range for MSW oscillations into sterile neutrinos is
\begin{equation}
P_{\rm ^7Be~sterile} = 0.006^{+0.25}_{-0.002}  .
\label{eq:globalebe7sterilesma}
\end{equation}
For vacuum oscillations, 
\begin{equation}
P_{\rm ^7Be~vacuum} = 0.30^{+0.42}_{-0.14}  ~.
\label{eq:globalebe7vac}
\end{equation}

With current knowledge, the allowed range of $P$ varies all the way
from $0.00$ to $0.72$, which emphasizes the importance of this
proposed measurement. 

\subsection{Global fits: energy-independent suppression}
\label{ss:globalconstant}

The preliminary results of the SuperKamiokande measurement of the
electron recoil energy spectrum provide additional evidence, beyond
that available from just the total rates discussed in 
Section~\ref{ss:constant}, 
regarding the hypothesis~\cite{nussinov76,harrison,foot,comforto,acker}  
for an energy-independent suppression of the solar 
neutrino fluxes (but see also the discussion of
Figure~\ref{fig:globalspectrum} in
Section~\ref{ss:dglobal})\footnote{Different physical mechanisms that
might lead to an energy-independent supression may predict different
zenith angle dependences for the detector event rates. Therefore, we
have not included information about the zenith-angle distribution in
the global fits for energy-independent supression.}. 
Assuming an energy-independent suppression of the
neutrino fluxes in the BP98 standard solar model~\cite{BP98}, 
 $\chi^2_{\rm min}$ for oscillations into active
neutrinos is 43 (17 d.o.f), i.e., this solution is disfavored at 99.95\%
C.L. . If the $^8$B neutrino flux is allowed to vary as a free parameter,
$\chi^2_{\rm min} = 42$ ( $16$
d.o.f), which is disfavored 
 at 99.96\% C.L. . Energy-independent oscillations
into sterile neutrinos are disfavored  at 99.998\% C.L. (minimum $\chi^2
= 50.3$) if one allows  an arbitrary $^8$B neutrino flux.
Maximally-mixed oscillations into active neutrinos 
which predict~\cite{harrison} a constant survival probability of  $5/9$ 
give $\chi^2_{\rm min} = 45$ 
($18$ d.o.f.),  which is disfavored at 99.96 \%
C.L. .For arbitrary $^8$B neutrino flux, 
 $\chi^2 = 44.5$ (17 d.o.f), which is disfavored  at 99.97\%
C.L. .

\section{DISCUSSIONS AND CONCLUSIONS}
\label{discussion}
We summarize and discuss in this section our principal conclusions.
In Sec.~\ref{ss:dindications}, we discuss  the indications that solar
neutrino experiments are suggesting new physics. We present 
in Sec.~\ref{ss:daveragerates}  the robust inferences
based only on the total event rates and  in Sec.~\ref{ss:ddaynight}  the
implications of the absence of a 
statistically significant zenith-angle dependence
for the SuperKamiokande event rates.  In 
Sec.~\ref{ss:dspectralshape},  we analyze 
the strong constraints imposed by the
electron recoil energy spectrum measured by SuperKamiokande.  Our
global analysis is described 
in Sec.~\ref{ss:dglobal}, where we
present  the implications of the combined constraints
from the total rates, the zenith-angle dependence, and the spectrum
shape. We summarize in Sec.~\ref{ss:dsummation} our overall view of
the solar neutrino situation.

For a given hypothesis, MSW or vacuum oscillations (active or sterile
neutrinos), we search numerically for the best-fit solution and quote
the C.L. for acceptance or exclusion based upon the relevant
$\chi^2_{\rm min}$ and the appropriate number of d.o.f. . The
acceptance levels found in this paper  seem somewhat higher than
reported by the SuperKamiokande Collaboration in their review talk at
Neutrino 98, but we think that this difference is largely due to the
fact that we have searched  for the best-fit solution rather than
test the acceptability of previously recognized solutions and that we
have used the BP98 rather than the BP95 predictions in analyzing the
total rates.  To the best of our knowledge, the present paper is the
first to determine global solutions that take account of the measured
total rates in all of the solar neutrino experiments, as well as the
SuperKamiokande zenith-angle distribution and the SuperKamiokande 
electron recoil energy spectrum.

As described below in Sect.~\ref{ss:dglobal},
the best-fit global MSW solution for active neutrinos is: $\Delta m^2  =  
5.4\times 10^{-6} {\rm eV}^2 ~,
\sin^22\theta  = 5.5\times 10^{-3}~ $ (and for sterile neutrinos:
$\Delta m^2  =  
4.0 \times 10^{-6} {\rm eV}^2 ~, \sin^22\theta  = 6.9\times 10^{-3}~ $).
For vacuum oscillations, the best-fit solution is:
$\Delta m^2  =  
6.5 \times 10^{-11} {\rm eV}^2 ~,
\sin^22\theta  = 0.75~ $ .

\subsection{Indications of new physics}
\label{ss:dindications}

The results from the first phase of the SuperKamiokande experiment
have strengthened the inference that new physics is required to
describe solar neutrino experiments. 

If we consider only the total rates in the
solar neutrino experiments, then there is no linear combination of the
undistorted neutrino energy spectra that can fit the available data at
the $3\sigma$ level (see the discussion in Sec.~\ref{sss:inclcno} and 
Sec.~\ref{sss:nocno}).  This result, whose physical basis is
described in Sec.~\ref{ss:discussionsm}, is independent of any
astrophysical arguments regarding the basic correctness of the solar
model.  
In particular, the $3\sigma$ discrepancy ignores the additional
evidence provided by 
helioseismological measurements, which agree to high  precision 
($0.1$\% r.m.s in sound speeds, see Ref.~\cite{reliable}) 
with the predictions of the standard
solar model. The Standard Solar
Model predictions are  inconsistent with the observed rates in solar neutrino
experiments at approximately the $20\sigma$ level (see Sec.~\ref{ss:ssms}). 

The data from the total rates alone indicate that the $\nu_e$ energy
spectrum from the sun is distorted, i.e., the survival probability
for electron type neutrinos to reach the earth is energy dependent.

If we impose the additional constraints from the measured SuperKamiokande
spectrum and Day-Night asymmetry as well as the total rates, then an
arbitrary linear combination of the $pp$, ${\rm ^7Be}$, ${\rm ^8B}$,
and CNO  neutrino
fluxes is ruled out at  the $3.5\sigma$ confidence level (see
Sec.~\ref{ss:arbglobal}). This minimum discrepancy again ignores 
all information about the solar model.

\subsection{Average event rates}
\label{ss:daveragerates}

The most robust results of the solar neutrino experiments so far are
the total observed rates, which are  summarized in
Table~\ref{datarates}.
We have therefore evaluated accurately the allowed regions of neutrino
parameters for either active or
sterile  MSW or vacuum neutrino oscillations, using the total rates as
the only constraints. If neutrino oscillations are indeed occurring,
a subset of the parameters that are consistent with the total rates
must also be consistent with the other measured quantities (electron
recoil energy spectrum and Day-Night asymmetry), a proposition that we
have also tested.

The principal results considering only the total rates are displayed in
Fig.~\ref{fig:mswrates}--Fig.~\ref{fig:vacvarb8}.
We have also given in Sec.~\ref{averagerates} the
best-fit neutrino oscillation parameters and mass differences for 
each scenario that we have discussed.

The most important change in the allowed range of neutrino parameters
compared to our previous work~\cite{brighter}
(which was prior to the announcement of the SuperKamiokande results)
is that the SMA solution is shifted to somewhat smaller mixing
angles,  
from the earlier value of 
$\sin^2 2\theta = 8.7 \times 10^{-3}$~\cite{brighter} 
to the current-fit value of 
$\sin^2 2\theta = 6.0 \times 10^{-3}$ (Eq.~\ref{SMApara}).
The main causes of this shift
are the smaller predicted ${\rm ^8B}$
neutrino flux (for the INT normalization) and the the lower
SuperKamiokande rate (somewhat lower than the Kamiokande rate). 
If the low energy cross section factor for the ${\rm
^7Be}(p,\gamma){\rm ^8B}$
reaction is treated as a free parameter, then the best-fit mixing
angle for the SMA solution decreases to an even smaller value of 
$\sin^2 2\theta = 3.5 \times 10^{-3}$ (Eq.~\ref{varSMApara}).
Further improvements in the most important nuclear cross sections are
unlikely to reduce significantly the size of the allowed neutrino
parameter regions (cf. Fig.~\ref{fig:mswratesS0}), although further moderate 
 shifts in the best-fit values may be anticipated.

The relatively modest  shift in mixing angles 
from $\sin^2 2\theta = 8.7 \times 10^{-3}$ 
to $\sin^2 2\theta = 6.0 \times 10^{-3}$ 
has the effect of reducing drastically the  predicted Day-Night
asymmetry from what was expected earlier~\cite{brighter}, i.e., $1.8$\% 
for the best-fit SMA solution, to a probably un-measurably small
$0.35$\% asymmetry for the current best-fit SMA solution (and is even
smaller for the best-fit global solution, see Sect.~\ref{ss:dglobal}
below).
The Day-Night asymmetry is obviously sensitive to details of the neutrino
parameter solutions.

Some theoretical models are already strongly disfavored by the
constraints of the total rates alone. For example, models in which the
suppression of electron type neutrinos is independent of energy are
excluded at the $99.8$\% C. L. or more, depending upon the precise
scenario considered (see the results described in Sec.~\ref{ss:constant}).

\subsection{Zenith-Angle dependence of rates}
\label{ss:ddaynight}

No statistically significant Day-Night asymmetry or zenith-angle
dependence of the solar neutrino event rate has been detected so far
by the SuperKamiokande 
experiment~\cite{superkamiokande300,superkamiokande374,superkamiokande504}.
The small-value of the observed Day-Night asymmetry excludes a large 
part  of the LMA region in neutrino parameter space that is allowed if only
the total solar neutrino rates are 
considered (see Fig.~\ref{fig:mswangular}).

\subsection{Electron recoil energy spectrum}
\label{ss:dspectralshape}

The electron recoil  energy distribution reported by
SuperKamiokande~\cite{superkamiokande504} is inconsistent 
with no distortion at about the $99$\% C.L. .  On the other hand, all
of the popular neutrino oscillation solutions determined from the total
rates only (SMA, LMA, and LOW MSW
solutions and vacuum oscillations) provide acceptable, although not
excellent, fits to the recoil energy 
spectrum (cf. Fig.~\ref{fig:linefit}).  
The observed
distortion of the spectrum, i.e., the overall slope parameter, is in
the direction predicted by the SMA solution.  However, as emphasized
by the SuperKamiokande Collaboration~\cite{superkamiokande504}, the fits
using the oscillation solutions for active neutrinos 
that are preferred by the total rates are not
particularly good. The last three points in the recoil spectrum are
somewhat higher than would have been expected  from an undistorted $^8$B
neutrino spectrum.

Figure~\ref{fig:spectrumMSWa} shows the regions of MSW parameter space
that are allowed 
for both active and sterile neutrinos by imposing only the 
constraint of consistency with the
electron recoil energy distribution.
The best-fit solutions for the spectrum only, marked by dark points in
Fig.~\ref{fig:spectrumMSWa}, do not fall within the allowed regions
determined by the global fit to all of the data (see
Sec.~\ref{ss:dglobal} below).
For vacuum oscillations, Fig.~\ref{fig:spectrumV} shows  the regions
allowed by the spectral data alone.

\subsection{Global analysis}
\label{ss:dglobal}

The combined constraints from the total rates, the zenith-angle dependence
(or the Day-Night Asymmetry), and the electron recoil energy
spectrum provide the most comprehensive test of neutrino oscillation
descriptions of solar neutrino experiments. 

The allowed solution space  for MSW oscillations, 
including the rates and the energy
spectrum,  is shown in Fig.~\ref{fig:global}a and the allowed solution
space 
including all three sets of constraints, rates, energy spectrum, and
zenith-angle dependence,  is  shown in Fig.~\ref{fig:global}b. The best-fit
solution, in both cases, is close to the best-fit  SMA
solution when only the total rates are considered. The best-fit
MSW solutions shown in  both Fig.~\ref{fig:global}a and Fig.~\ref{fig:global}b
are not particularly good, but are acceptable at the $7$\% C. L. .
When all three sets of constraints are imposed,
the SMA solution space is reduced somewhat in size compared
to what is allowed if only the total rates are considered.
For sterile neutrinos, the SMA solution is acceptable at the $8$\%
C.L. ; the allowed region, which is similar to what is obtained for
active neutrinos,  is shown in Fig.~\ref{fig:globalMSWst}  .
The  LMA and LOW solutions are excluded for active and sterile
neutrinos  at the $99$\% C. L. when
all three constraints, rates, spectrum, and Day-Night asymmetry, are
included.

The global fits of MSW solutions to all the available data predict a
Day-Night asymmetry, defined by Eq.~(\ref{eq:daynight}), 
 for the SuperKamiokande experiment (total energy
$> 6.5 $ MeV) that lies in the range $-0.0048 < A < 0.025$ for
active neutrinos and $-0.0031 < A < 0.007$ for sterile neutrinos.  The
best-fit solutions predict an asymmetry whose absolute value is less
than $0.1$\% , un-measurably small,  for both active and sterile neutrinos.

For vacuum oscillations, Fig.~\ref{fig:globalV} shows the allowed
solution space when the rates, energy spectrum, and zenith-angle
dependence are all imposed. The best-fit vacuum solution is acceptable
at the $6$\% C. L. . The inclusion of the Day-Night and the spectral
constraints reduces considerably the parameter regions for vacuum
oscillations (cf. Fig.~\ref{fig:vacrates}).

Figure~\ref{fig:globalspectrum} shows how well (or poorly) the
calculated global oscillation solutions fit the observed
SuperKamiokande~\cite{superkamiokande504} electron recoil energy
spectrum. Each of the three panels shows for a different global
oscillation solution (described in Section~\ref{global}) the
ratio of the number of electrons in a given energy bin to the number
expected using the undistorted neutrino energy spectrum~\cite{balisi}
and the electroweak neutrino-electron scattering cross sections with
radiative corrections~\cite{sirlin}. 
The data shown were presented by the SuperKamiokande collaboration at
Neutrino98~\cite{superkamiokande504}. In computing the predictions for
the different global oscillation solutions, we included the reported
SuperKamiokande energy resolution and trigger efficiency function.
All three of the best-fit global solutions fall well below the
measured Ratio in the three highest energy bins. The possibility that
this discrepancy at high energies 
might be due to a larger-than-expected cross section
for the production of  $hep$ neutrinos is discussed extensively by
Bahcall and Krastev~\cite{hep98}. The global best-fit vacuum
oscillation solution shows a slight upturn at the lowest energies
that might be detectable when SuperKamiokande obtains data with a
threshold that extends to $5$ MeV .

The neutrino oscillation predictions for the neutrino-electron
scattering rate of 
the $0.86$ MeV ${\rm ^7Be}$  line, which will be measured by the BOREXINO
neutrino experiment, are
robust.  Very similar predictions are obtained using only the total rates
as a constraint and using the presently available data on the spectral
shape and Day-Night asymmetry as well as the total rates.  The results
are summarized in Eq.~(\ref{eq:globalbe7sma}),  
Eq.~(\ref{eq:globalbe7sterilesma}), and Eq.~(\ref{eq:globalbe7vac}).

We have also evaluated the globally allowed range of the $\nu_e$
survival probability which could be compared with future measurements
of the $\nu_e$ $^7$Be neutrino flux, in a charged current (LENS)
experiment. The results are given in Eq.~(\ref{eq:globalebe7sma}),  
Eq.~(\ref{eq:globalebe7sterilesma}), and Eq.~(\ref{eq:globalebe7vac}).

\subsection{Summation}
\label{ss:dsummation}

Different neutrino oscillation scenarios, including SMA MSW conversion
to either active or sterile neutrinos and vacuum conversion to active
neutrinos, give similar quality descriptions of the solar neutrino 
data, acceptable at the 
level $6$\%-$8$\%.  Moreover, with the strong evidence for atmospheric
neutrino oscillations than is now available, 
an oscillation solution of the solar neutrino
problem seems even more probable than before. In addition, any
description of solar neutrinos that does not include some new physics
that causes an energy dependence in the $\nu_e$ survival probability
is strongly disfavored by the combined data from solar neutrino
experiments, 
while standard solar model calculations 
accurately  predict the helioseismologically 
measured sound velocities in the sun.

However, with the existing data it is not possible to determine which
kind of transition(s) solar neutrinos undergo. Fits of neutrino
oscillation scenarios to just the total event rates in the chlorine,
Kamiokande, GALLEX, SAGE, and SuperKamiokande experiments, which may
be the most robust currently available experimental information, suggest
neutrino parameters in which the Day-Night asymmetry and the spectral
distortion are  difficult to measure.  The situation may improve
significantly when data from the lower-energy spectral bins of
SuperKamiokande, at $5.5$ MeV and $6.0$ MeV, are available.  Crucial
information will also be provided by the future SNO~\cite{McD94}, 
BOREXINO~\cite{borexino}, and GNO~\cite{gno} experiments. 

We suspect that the unique description of solar neutrino phenomena
will require global analysis of all of the
available data.  We hope that this paper is a useful step along the
path toward a complete solution of the solar neutrino problems.

\acknowledgments
We are grateful to the chlorine, Kamiokande, GALLEX, SAGE, and
SuperKamiokande collaborations for the superb data that made possible
this analysis. We are especially grateful to the SuperKamiokande
collaboration for making available preliminary data at Neutrino 98, in
addition to the data in their formal paper~\cite{superkamiokande374}, 
which permitted more stringent inferences.
JNB  acknowledges support from NSF grant No. PHY95-13835 and
stimulating discussions with M. Cribier and R. Raghavan regarding the 
proposed LENS
experiment. 
PIK acknowledges support from  NSF grant No. PHY95-13835  and
NSF grant No. PHY-9605140.  
AS thanks the Institute for Advanced Study for hospitality during a
visit in which this work was initiated.

\appendix
\section*{Analysis details}

We describe here the methods used in our statistical analysis of the
solar neutrino data.

\subsection{Rates only}

In the analysis which includes only the event rates in the different
detectors, we use the procedure described in detail in Ref.~\cite{fogli95}.
The $\chi^2$ for the combined fit is defined as:

\begin{equation}
\chi^2({\rm Rates}) = \Sigma_{i,j = 1, 4} (R^{\rm th}_i - R^{\rm exp}_i)
V^{-1}_{i,j} ( R^{\rm th}_j - R^{\rm exp}_j ).\hskip 2cm
\end{equation}
Here $R^{\rm th}_i$ is the theoretically predicted event rate in the
$i$-th detector (chlorine, SAGE, GALLEX and SuperKamiokande), which
takes into account the electron neutrino survival probabilities,
$R^{\rm exp}_i$ is the corresponding experimentally measured event rate,
and $V_{i,j}$ is the error matrix which is a function of the
theoretical uncertainties (nuclear cross-sections, age, luminosity and
heavy element abundances in the standard solar model) as well as the
experimental statistical and systematic errors from each experiment
 (see Table~\ref{datarates}). The theoretical uncertainties are determined  in
BP98~\cite{BP98}. 
Following BP98, we have included theoretical uncertainties due to the
diffusion of elements in the sun and to  the electron capture cross
section on ${\rm ^7Be}$, not previously included as uncertainties in
neutrino oscillation studies.  A detailed description of how the individual
uncertainties are calculated can be found at
http://www.sns.ias.edu/$\sim$jnb/SNdata in the menu item `Solar neutrino
rates, fluxes, and uncertainties.'
Important correlations exist between the neutrino
fluxes that
must be included correctly in the evaluation of the error matrix or
the calculated allowed regions will be incorrectly reduced in size.

\subsection{Recoil electron spectrum}

Next we describe the analysis of the recoil electron spectrum in
SuperKamiokande which includes a computation of the following $\chi^2$
function:

\begin{equation}
\chi^2(\rm Spectrum) = \Sigma_{i,j = 1, 16} 
(\alpha S^{\rm th}_i - S^{\rm exp}_i)
W^{-1}_{i,j} ( \alpha S^{\rm th}_j - S^{\rm exp}_j ).\hskip 2cm 
\label{eqa:chisqspec}
\end{equation}
Here $S^{\rm th}_i$ is the theoretically predicted event rate for the
$i$-th energy bin in SuperKamiokande and $S^{\rm exp}$ is the
corresponding experimentally measured event rate.  
In this paper, we use the measured electron recoil 
energy spectrum data presented by the SuperKamiokande
Collaboration~\cite{superkamiokande504} 
at Neutrino 98 and the recently determined undistorted
spectrum~\cite{balisi}. The statistical as well
as the systematic experimental errors are included. Since the latter
are fully correlated, we assume a correlation coefficient of 1 between
each pair of bins.  Neglect of  these correlations would lead  to an 
appreciable
increase of the regions that is excluded. The entries in the
covariance matrix are:

\begin{equation}
W_{i,j} = \sigma^{\rm stat}_i\sigma^{\rm stat}_j \delta_{i,j} +
\sigma^{\rm syst}_i\sigma^{\rm syst}_j ~.
\label{eqa:covariance}
\end{equation}

The coefficient $\alpha$ in Eq.~(\ref{eqa:chisqspec})
is an overall normalization coefficient and
is varied as a free parameter independent of $\Delta m^2$ and
$\sin^22\theta$. This variation 
 reflects the fact that  we are interested here in a
test of the shape of the measured spectrum and not in the overall
event rate in SuperKamiokande. 

We calculate the theoretically expected event rates in the individual
bins including the quoted energy resolution and trigger efficiency
function in SuperKamiokande. We neglect the uncertainty in the energy
normalization since, according to the SuperKamiokande Collaboration,
it is less than 1\%. 

\subsection{Day-Night asymmetry and zenith angular dependence}

The Day-Night asymmetry measured by
SuperKamiokande~\cite{superkamiokande504}  is not
significantly different from zero, which limits the allowed
neutrino oscillation parameters. Assuming $3\sigma$ errors and
combining the statistical and systematic errors in quadrature, the
asymmetry $A_{\rm exp}$, defined by Eq.~(\ref{eq:daynight}), is

\begin{equation}
 -0.096 < A_{\rm exp} < 0.050 .
\label{eqa:aexp}
\end{equation}

The lower limit in Eq.~(\ref{eqa:aexp}) 
is not especially useful since we have found by explicit
calculation  that  asymmetries predicted 
by MSW parameters in the entire range 
$ 10^{-9} {\rm eV}^2 < \Delta m^2 <  10^{-3} {\rm eV}^2 $ and 
$10^{-4} < \sin^22\theta <  1.0 $,
never take on very large negative values 
 $(A > -0.005)$. However, the upper limit is   exceeded for
certain choices of previously-allowed  oscillation 
parameters;  these parameters  are therefore excluded by the SuperKamiokande
measurement.

Since the significant theoretical uncertainties cancel in the ratio, 
we define
$\chi^2$ for the Day-Night  asymmetry as:

\begin{equation}
\chi^2(D/N) = (A_{\rm th} - A_{\rm exp})/ \Delta A_{\rm exp} ~, \hskip 2cm
\label{eqa:daynightchi}
\end{equation}
where $A_{\rm th}$ is the theoretically predicted asymmetry due to
neutrino regeneration in the earth (zero for vacuum oscillations)
and $\Delta A_{\rm exp} = 0.024$ is the
combined statistical plus systematic error given by the
SuperKamiokande Collaboration.

Figure~\ref{fig:mswangular} and 
Figure~\ref{fig:mswangularsterile}  show the excluded regions for active
and sterile neutrinos, respectively. 
Note that the relatively large uncertainties in the total ${\rm ^8B}$
neutrino flux cancel in forming the asymmetry ratio and therefore do
not affect the excluded regions. Also, we have verified that at the
present level of accuracy the excluded regions are not affected
significantly by the uncertainty in the undistorted ${\rm ^8B}$
neutrino spectrum.

The measured zenith angular distribution  provides another constraint
on neutrino oscillation parameters. We have calculated the predicted
zenith angular distribution of events for many points in the
$\sin^22\theta$ - $\Delta m^2$ plane and have compared the predicted
distributions with the binned zenith-angle distribution 
 presented~\cite{superkamiokande504} by the SuperKamiokande
Collaboration at the Neutrino 98 conference.

The $\chi^2$ in this case is defined by

\begin{equation}
\chi^2(\rm Zenith) = \Sigma_{i,j = 1, 10} (\alpha Z^{\rm th}_i -
Z^{\rm exp}_i)
U^{-1}_{i,j} ( \alpha Z^{\rm th}_j - Z^{\rm exp}_j ).\hskip 2cm
\label{eqa:chizenith}
\end{equation}

The predicted distribution  for the zenith angle dependence 
was calculated  by
assuming uninterrupted operation of the SuperKamiokande detector during 504
days  starting from April 1, 1996. This assumption is not precisely correct
since the detector has been occasionally
shut down for calibration and  maintenance. However, since
these interruptions are a small fraction of the  data taking
period, our calculation is not seriously degraded by the lack of
published information about the detailed operating schedule.

The SuperKamiokande Collaboration can perform a unique public service
by testing the sensitivity of their inferences 
based upon the zenith
angle dependence to the assumptions regarding the operating
schedule. They can test at what quantitative level the small effects
due to the precise operating schedule affect the conclusions regarding
neutrino parameters.

The matrix $U$ that appears in Eq.~(\ref{eqa:chizenith}) 
 is assumed, in our calculations, to be  diagonal. 
Correlations between the 10 bins of the
zenith angular distribution  might conceivably arise from systematic effects
(for example, electrons falling in different bins being detected by the same
photomultiplier tubes). Although these correlations are expected to be
 small, it would be very useful if the SuperKamiokande Collaboration
 were to publish their estimates of any correlations that cause $U$ to
 be non-diagonal.

\subsection{Straight line fit}

As discussed in Sec.~\ref{ss:distorted}, the ratio, $R$,  of the measured to
  the expected spectral energy distribution can be 
  fit by a straight line, as in Eq.~(\ref{eq:ratioR}.
For an undistorted spectrum, $R$ is a constant independent of 
energy.

The $\chi^2$ in this case is defined as a sum over the $16$ energy bins
of the electron spectrum:

\begin{equation}
\chi^2(\rm Line) = \Sigma_{i,j = 1, 16} (\alpha L^{\rm th}_i - S^{\rm exp}_i)
W^{-1}_{i,j} ( \alpha L^{\rm th}_j - S^{\rm exp}_j ) ~,
\end{equation}
where $L^{\rm th}_i = R_0 + S_0 (E_{e,i} - E_0)$, $E_0 = 10$ MeV is a
conveniently-chosen energy, $E_{e,i}$ is the total electron energy in
the i-th bin (we use the energy in the middle of each energy bin), and
$S_0$ and $R_0$ are arbitrary parameters which are varied until the
minimum $\chi^2$ is found. The covariance matrix $W$ is the same as the
covariance matrix  defined in Eq. ~(\ref{eqa:chisqspec}). 

\subsection{Global fits}

In Sec.~\ref{global}, we combine in different ways 
the constraints from the measured
total rates, the spectrum shape, and the Day-Night asymmetry (or
zenith-angle dependence). 
In each case, we add the $\chi^2$ for each data set (
rates, energy spectrum, and Day-Night asymmetry or zenith-angle
distribution).  The individual 
$\chi^2$ are defined in previous sections of this
appendix and are treated as independent in the global fits.
Thus in Sec.~\ref{global} the effective $\chi^2$ are always the sum
of two or three independent terms.

\begin{table}
\caption[]{Solar neutrino data used in the analysis. The experimental
results are given in SNU for all of the experiments except Kamiokande
(and SuperKamiokande),
for which the result is expressed as the measured ${\rm ^8B}$ flux
above 7.5 MeV (6.5 MeV) 
in units of ${\rm 10^6 cm^{-2}s^{-1}}$ at the earth. The
ratios of the measured values to the corresponding predictions in the
 Bahcall-Pinsonneault standard solar model (BP98) of 
Ref.~\protect\cite{BP98} are also given. The
INT normalization (or, in parentheses, the 
CIT normalization) is used in these calculations.
Only experimental errors are included in the column labeled
Result/Theory.  The
results cited for the Kamiokande and SuperKamiokande experiments
assume that the shape of the ${\rm ^8B}$ neutrino spectrum is not
affected by physics beyond the standard electroweak
model.\label{datarates} }
\begin{tabular}{l c c c c c c}
Experiment & Result &\multicolumn{2}{c}{Theory}
&\multicolumn{2}{c}{Result/Theory}&Reference\\
\hline

Homestake & $2.56 \pm 0.16 \pm 0.16$ &
$7.7^{+1.2}_{-1.0}$& $(8.8^{+1.4}_{-1.1})$& $0.33 \pm 0.029$&
(0.29)&\ \ \cite{homestake}\\

Kamiokande & $2.80 ~\pm 0.19 ~\pm 0.33$ & $5.15 ~^{+1.0}_{-0.7}$ 
& $(6.1^{+1.1}_{-0.9})$& $0.54 \pm 0.07$&(0.46)&\ \ \cite{kamiokande}\\ 

GALLEX & $77.5 ~\pm 6.2 ~{}^{+4.3}_{-4.7}$ & $129^{+8}_{-6}$&$(131^{+9}_{-7})$&
$0.60 \pm 0.06$&(0.59)&\ \ \cite{GALLEX}\\

SAGE & $66.6 ~{}^{+7.8}_{-8.1}$ & $129^{+8}_{-6}$&$(131^{+9}_{-7})$&
$0.52 \pm 0.06$&$(0.51)$&\ \ \cite{SAGE} \\

SuperKamiokande & $2.44 ~\pm {+0.05} ~{}^{+0.09}_{-0.07}$ 
& $5.15 ~^{+1.0}_{-0.7}$&$(6.1^{+1.1}_{-0.9})$& $0.474 \pm 0.020$ 
&(0.39)&\ \ \cite{superkamiokande504}\\ 

\end{tabular}
\end{table}

\begin{table}
\caption[]{Minimum $\chi^2$ for different neutrino oscillation
solutions of the solar neutrino problems The reference solar model
corresponds to the INT normalization (or, in parentheses, the 
CIT normalization) of $S_{17}(0)$.
Results are given for oscillations into either either active or
sterile neutrinos.
\label{tab:minchi} }
\begin{tabular}{l c c c c c}

Solution & & MSW & & Vacuum Oscillations \\
\hline
 & SMA & LMA & LOW & \\
\hline
active & 1.7 (2.8) & 4.3 (1.9) & 7.4 (8.2) & 4.3 (2.7)\\
\hline
sterile& 1.7 (1.6)& 19.0 (17.0) & 17.0 (16.0) & 12.0 (11.2) \\

\end{tabular}
\end{table}

\begin{figure} 
\centerline{\epsfxsize=5.5in\epsffile{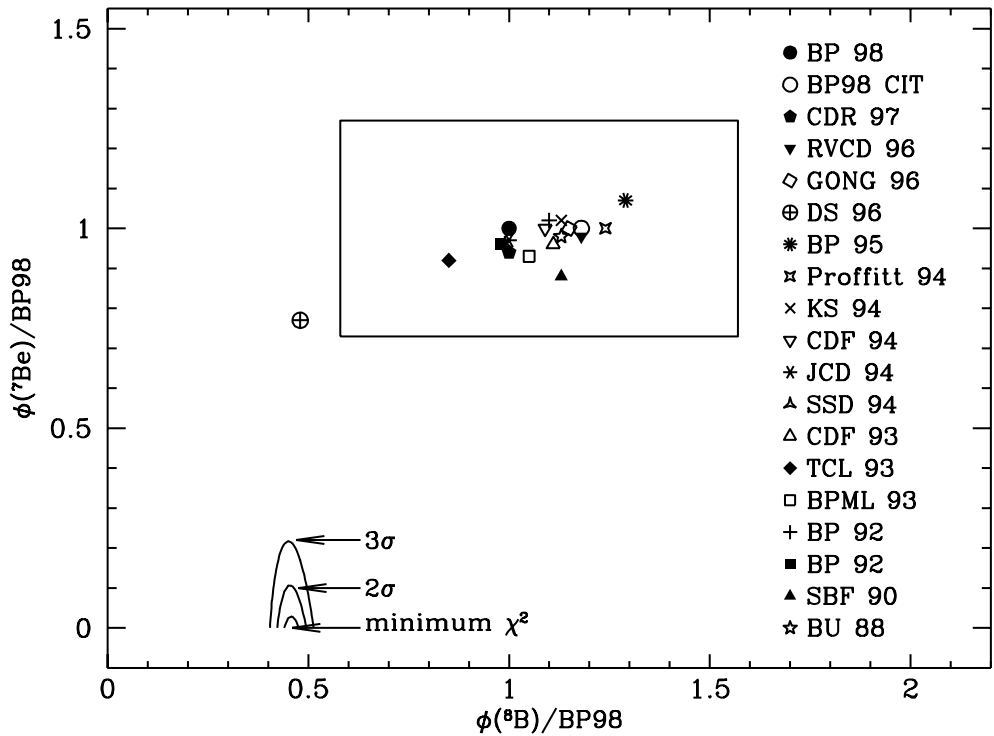}}
\vglue.2in
\caption[]{Predictions of standard solar models since 1988.
The figure
shows the predictions of $19$ standard solar models in the
plane defined by the ${\rm ^7Be}$ and ${\rm ^8B}$ neutrino fluxes. 
The abbreviations that are used in the figure to identify different
solar models are defined in the bibliographical item, Ref.~\cite{models}.
We include all standard solar models with which we are familiar that
were published in refereed journals
in the decade 1988-1998.
All of the
fluxes are normalized to the predictions of the 
Bahcall-Pinsonneault 98 solar model, BP98~\cite{BP98}.
The rectangular error box defines the $3\sigma$ error range of the
BP98 fluxes. The best-fit ${\rm ^7Be}$ neutrino flux is negative. At the
$99$\% C.L., there is no solution with all positive neutrino fluxes
if the fluxes of CNO neutrinos are arbitrarily set equal to
zero. There is no solution at the $99.9$\% C.L. if the CNO neutrinos
are fixed at their standard solar model values.
All of the standard model solutions lie  far from the best-fit
solution, even far from the $3\sigma$ contour.}
\label{fig:independent}
\end{figure}

\begin{figure} 
\epsfxsize=6.5in\epsffile{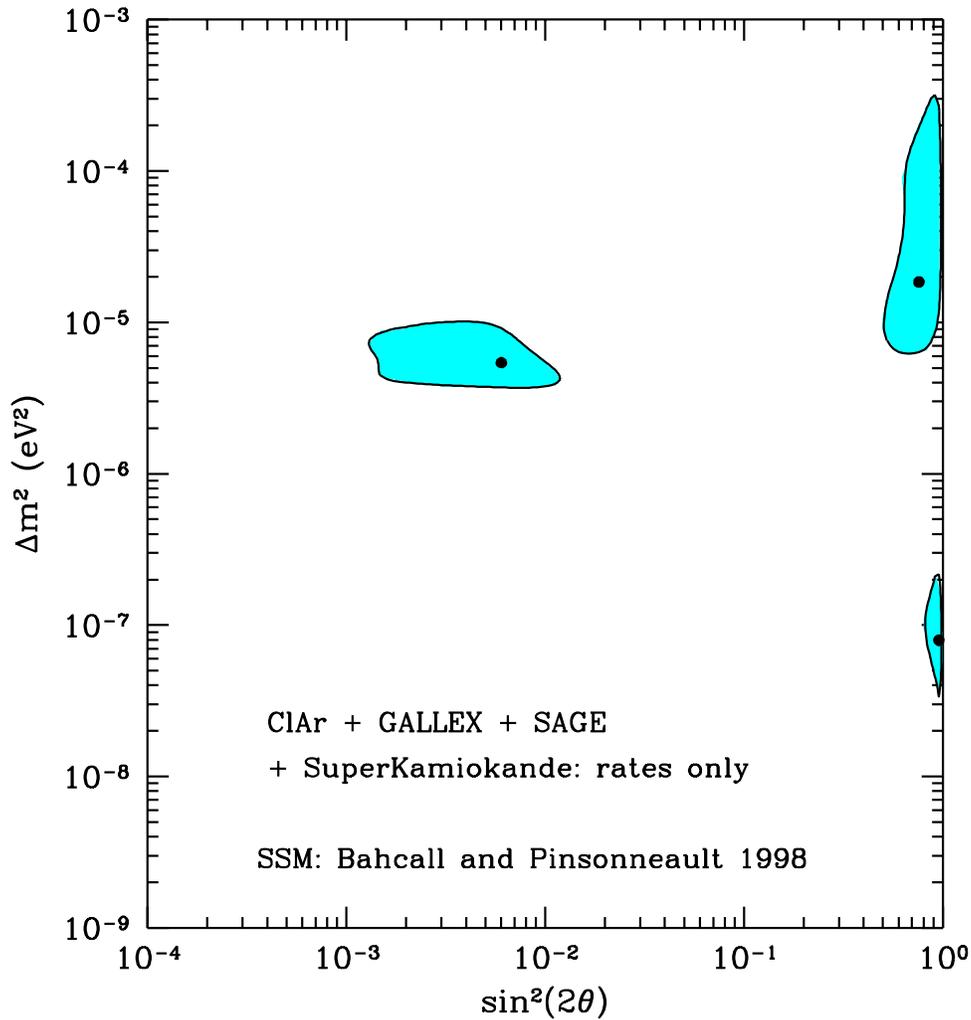}
\caption[]{MSW solutions: rates only. 
The figure shows the allowed  regions (99\% C.L.) in $\Delta m^2$ ---
$\sin^22\theta$ parameter space for the MSW solution. 
The best fit points are indicated by dark circles.
Only the total event rates in the chlorine, SuperKamiokande, GALLEX,
and SAGE experiments 
are considered; the solar neutrino data are
summarized in Table~\ref{datarates}.
The ${\rm ^8B}$
neutrino flux corresponds to the ``INT normalization'' of ${\rm
S}_{17}(0)$. The neutrino transitions in the sun are assumed to be
between active neutrinos ($\nu_e\rightarrow\nu_\mu$ or
$\nu_e\rightarrow\nu_\tau$).   }
\label{fig:mswrates}
\end{figure}

\vbox{
\begin{figure} 
\centerline{\epsfxsize=4in\epsffile{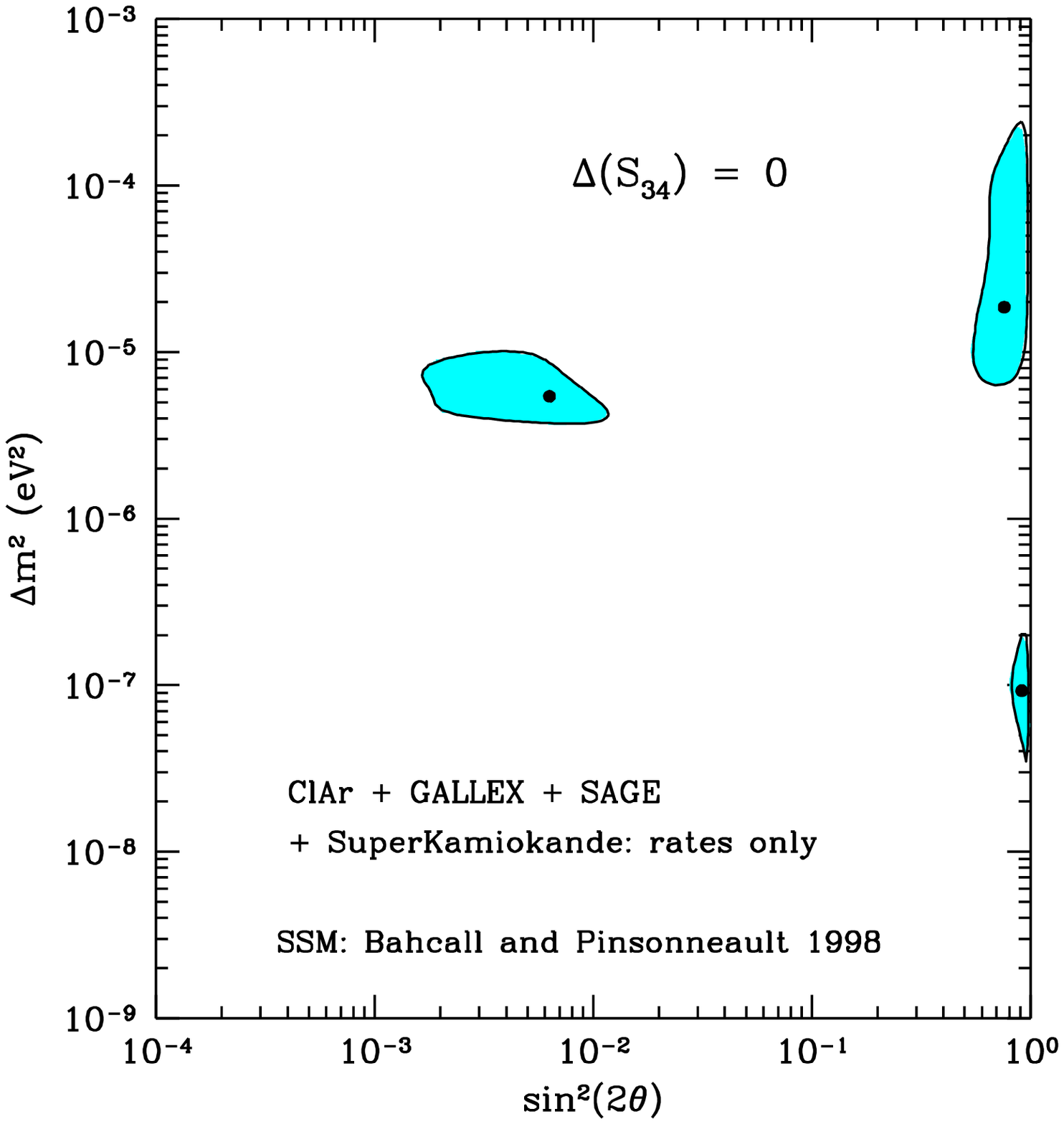}}
\vglue-.5in
\centerline{\epsfxsize=4in\epsffile{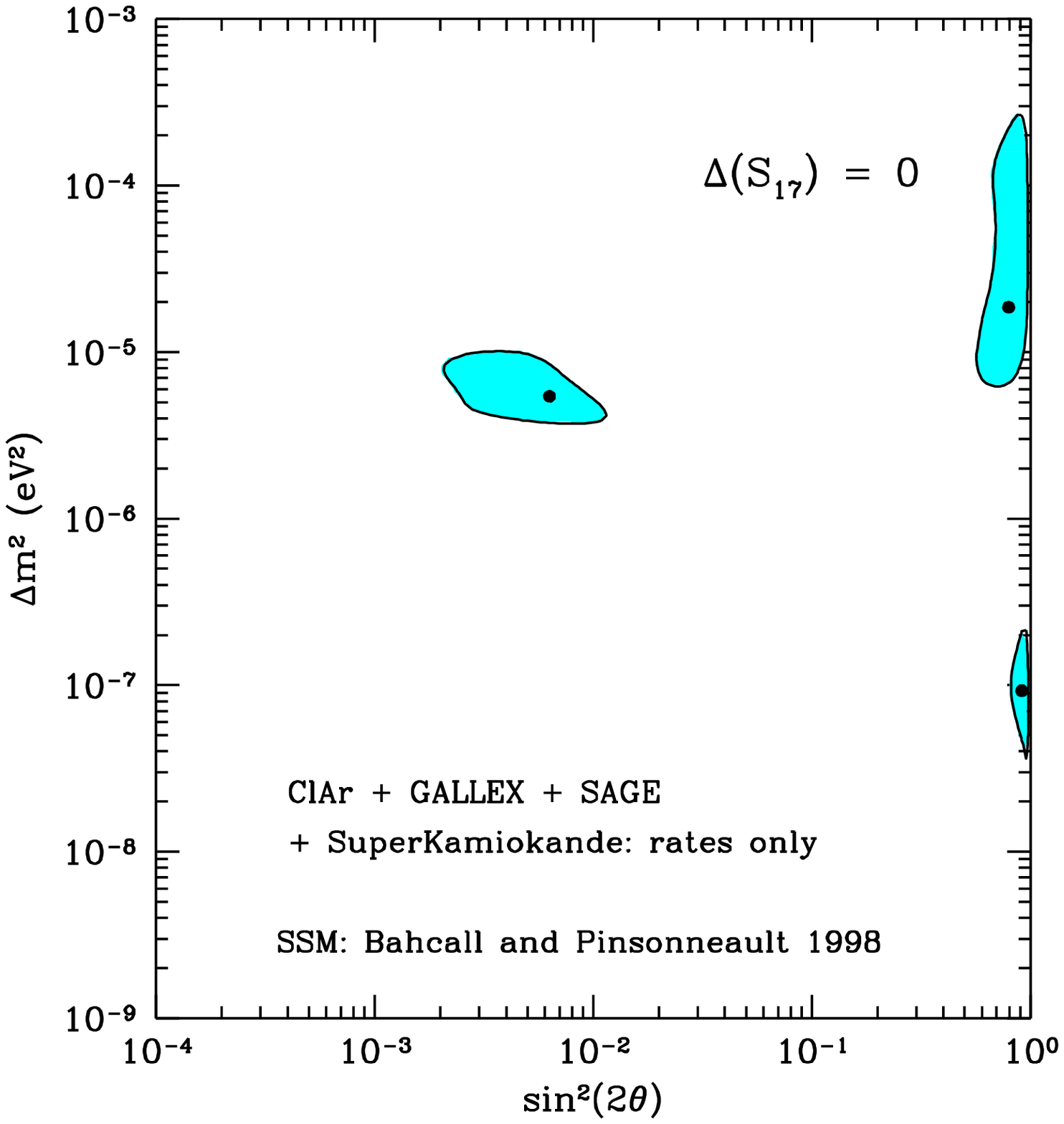}}
\caption[]{Effect of uncertainties in $S_{34}$ and $S_{17}$
on the allowed MSW solution space.
The panels are the same as in Fig.~\ref{fig:mswrates} except that
in the top panel the uncertainty in $S_{34}$ is set equal to zero
and in the lower panel the uncertainty in $S_{17}$ is set equal
to zero. There is no large reduction in the  MSW allowed regions if 
either $S_{34}$ or $S_{17}$ is presumed to be negligible.}
\label{fig:mswratesS0}
\end{figure}
}

\begin{figure} 
\epsfxsize=6.5in\epsffile{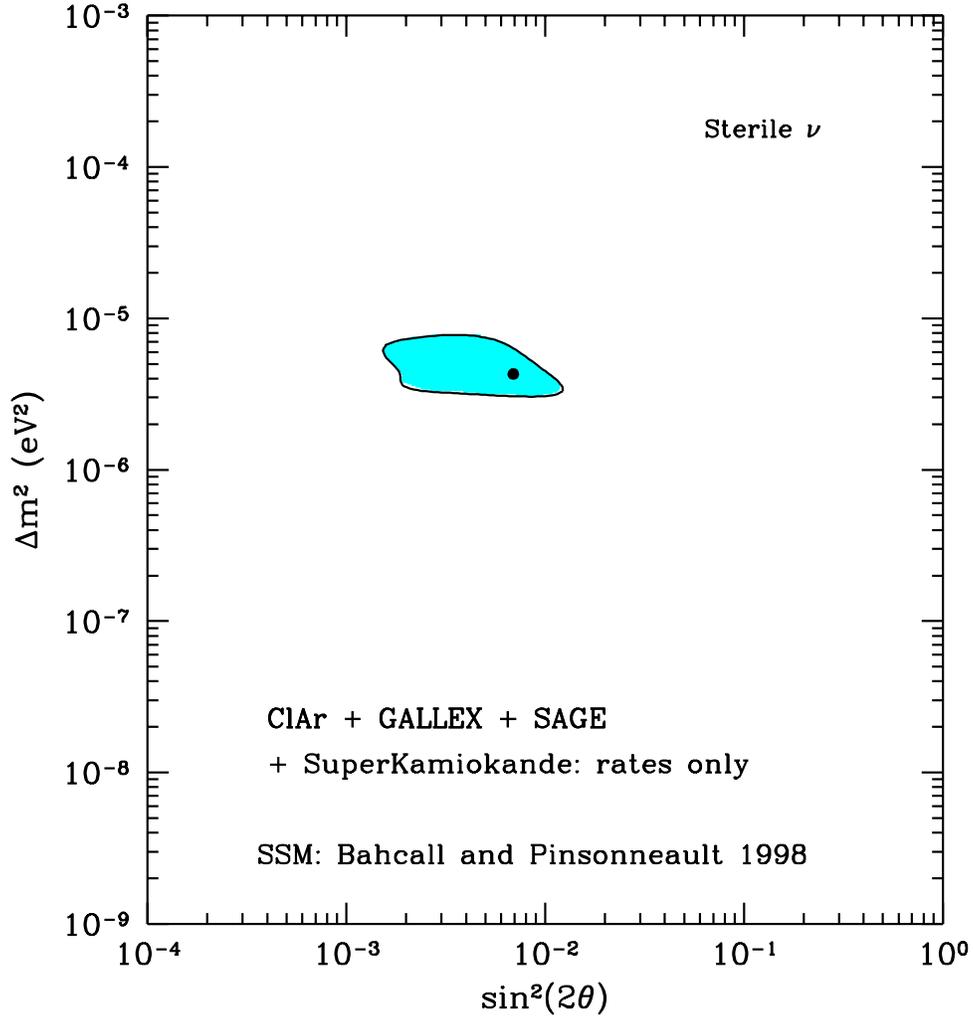}
\caption[]{Sterile neutrinos: rates only. 
The figure shows the  regions  in $\Delta m^2$ ---
$\sin^22\theta$ parameter space that the total rates in the chlorine,
SuperKamiokande, GALLEX, and SAGE experiments allow 
for MSW oscillations between an electron type
neutrino and a
sterile neutrino ($\nu_e\rightarrow\nu_s$).  Other conditions are the
same as for Figure~\ref{fig:mswrates}. }
\label{fig:mswratesst}
\end{figure}

\begin{figure} 
\epsfxsize=6.5in\epsffile{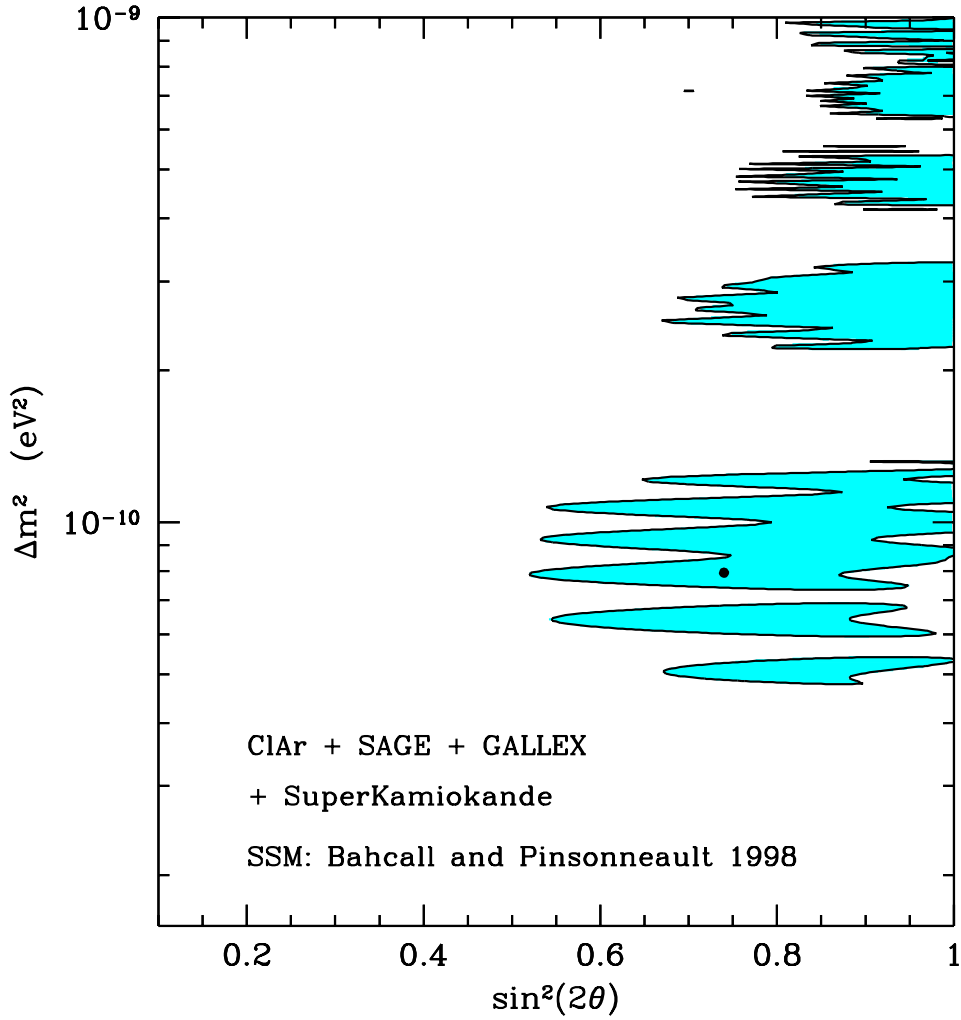}
\caption[]{Vacuum oscillations: rates only.
The figure shows the regions  in $\Delta m^2$ ---
$\sin^22\theta$ parameter space  that the total rates in the chlorine,
SuperKamiokande, GALLEX, and SAGE experiments allow for  vacuum neutrino 
oscillations between active
neutrinos.  The best-fit point is indicated by a dark circle.
The  experimental rates are summarized in Table
\ref{datarates}.  }
\label{fig:vacrates}
\end{figure}

\begin{figure} 
\epsfxsize=6.5in\epsffile{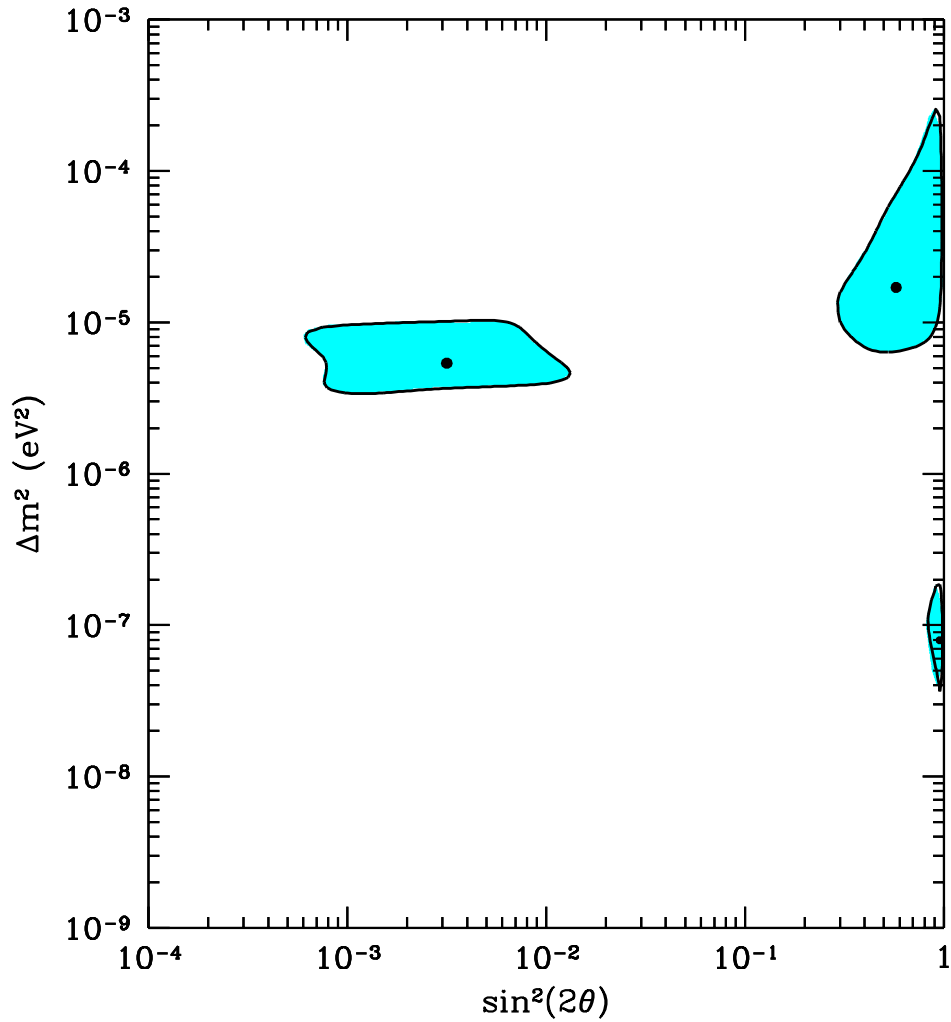}
\caption[]{ Variable ${\rm ^8B}$ flux and MSW solutions. The figure
shows the allowed regions at 99\%
C.L. in the $\Delta m^2$ -- $\sin^22\theta$ plane
for the MSW solutions with  an arbitrary ${\rm ^8B}$  neutrino flux (treated as
a free parameter).} 
\label{fig:mswvarb8}
\end{figure}

\begin{figure} 
\epsfxsize=6.5in\epsffile{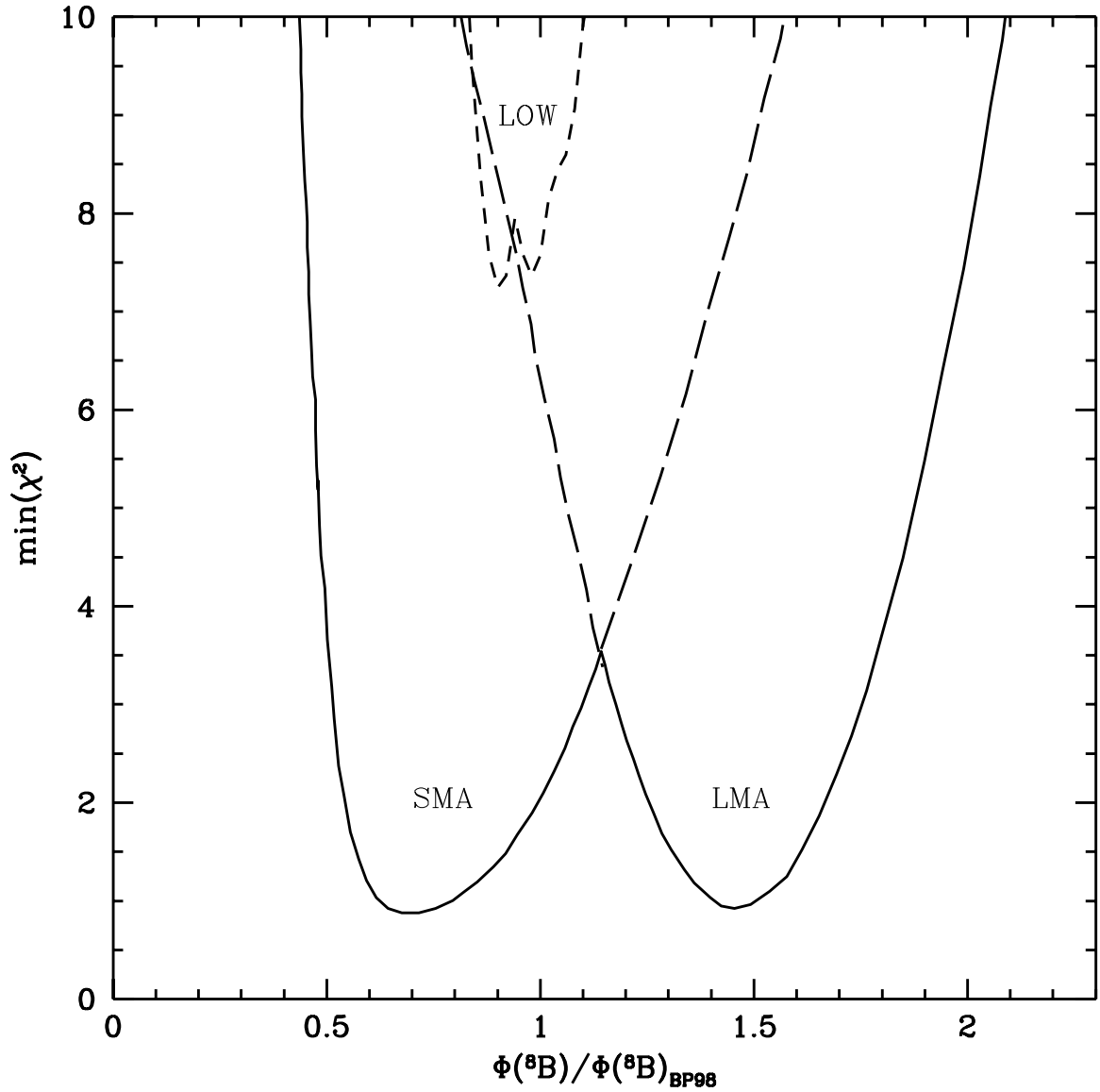}
\caption[]{Minimum $\chi^2$ (MSW solution) as a function of the boron
neutrino flux. The oscillations are between active neutrinos
($\nu_e\rightarrow\nu_\mu$ or $\nu_e\rightarrow\nu_\tau$). The
reference boron flux in the BP98 solar 
model corresponds to INT normalization of
$S_{17}(0)$. }
\label{fig:minchivarb8}
\end{figure}

\begin{figure} 
\epsfxsize=6.5in\epsffile{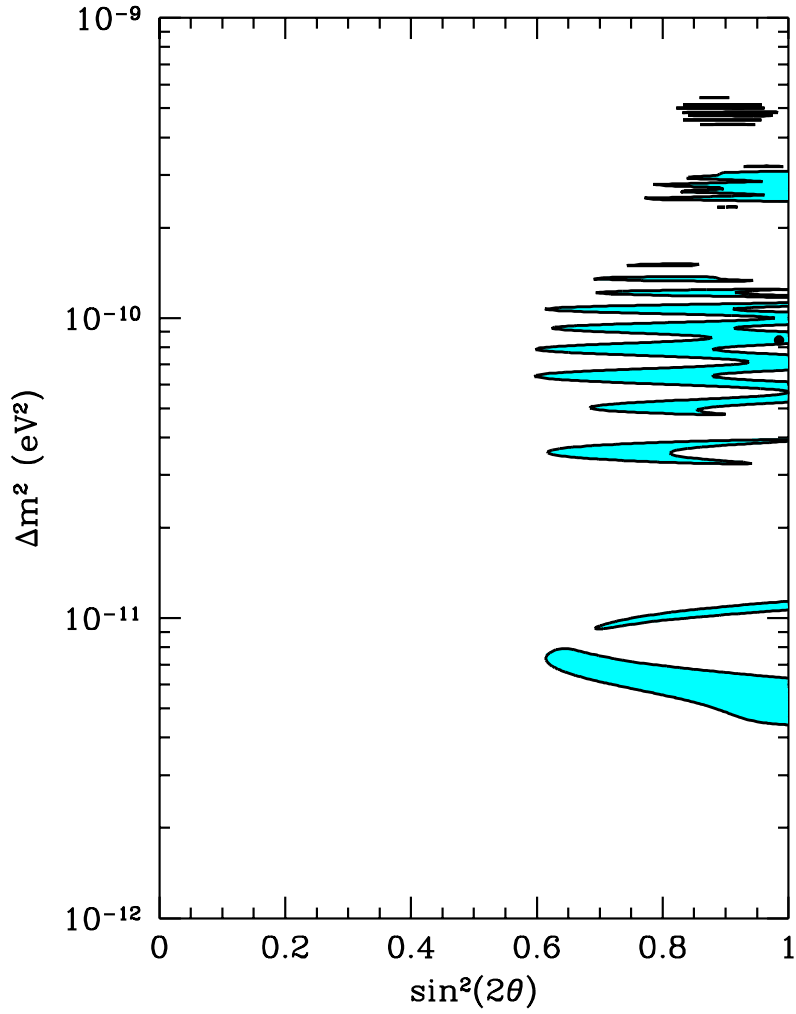}
\caption[]{Allowed regions in $\Delta m^2$ --
$\sin^22\theta$ parameter space for vacuum oscillations with an
arbitrary ${\rm ^8B}$ neutrino flux. The oscillations are assumed to occur
between active neutrinos.}
\label{fig:vacvarb8}
\end{figure}

\vbox{
\begin{figure}
\epsfxsize=6in\epsffile{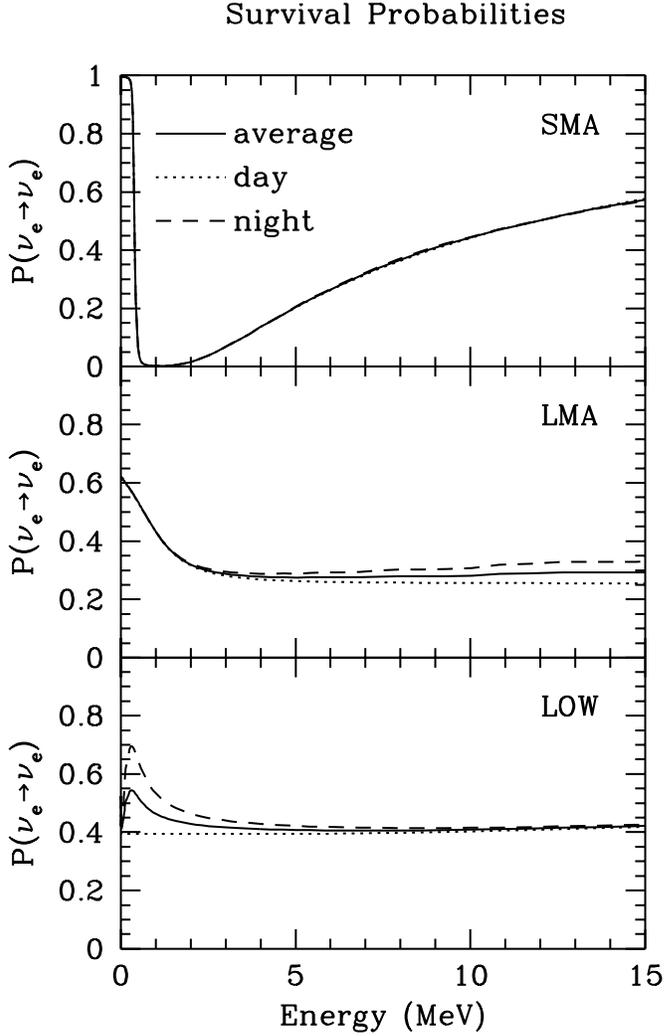}
\caption[]{Survival probabilities for MSW solutions.
The figure presents the yearly-averaged 
survival probabilities for an electron neutrino that is  created
in the sun to remain an electron neutrino 
 upon arrival at the SuperKamiokande
detector.  The best-fit MSW solutions including regeneration in the
earth are described in
Sec.~\ref{averagerates}.
The full line refers to 
the average
survival probabilities computed  taking
into account regeneration in the earth and the dotted line refers to
calculations for the daytime 
that do not include regeneration. 
The dashed line includes regeneration at night.
There are only slight
differences between the computed regeneration probabilities for
the detectors located at the positions of Super-Kamiokande, 
SNO and the Gran Sasso
Underground Laboratory (see Ref.~\cite{brighter}).
\label{fig:survival}}
\end{figure}
}

\vbox{
\begin{figure} 
\centerline{\epsfxsize=3.75in\epsffile{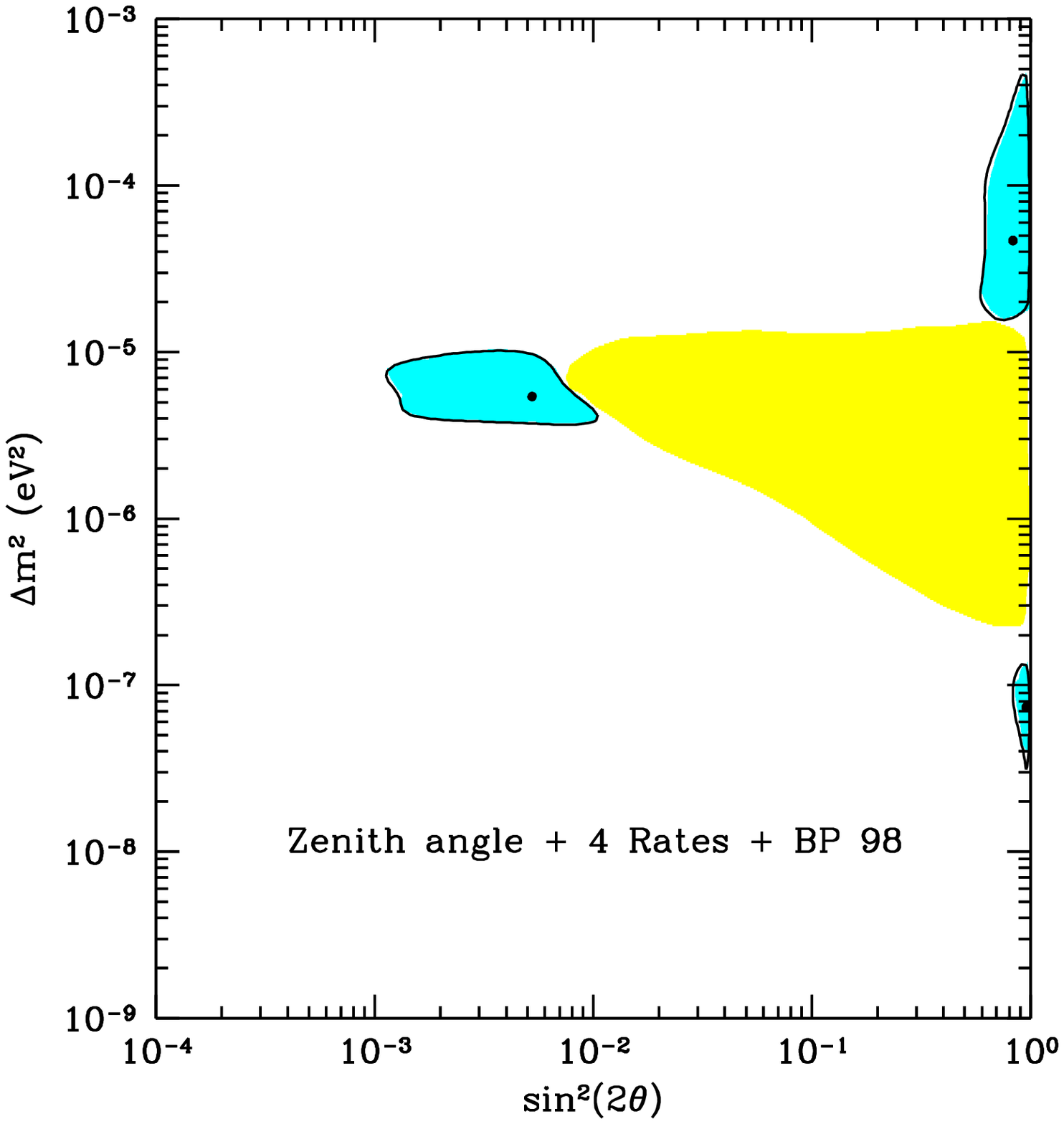}}
\vglue-.5in
\centerline{\epsfxsize=3.75in\epsffile{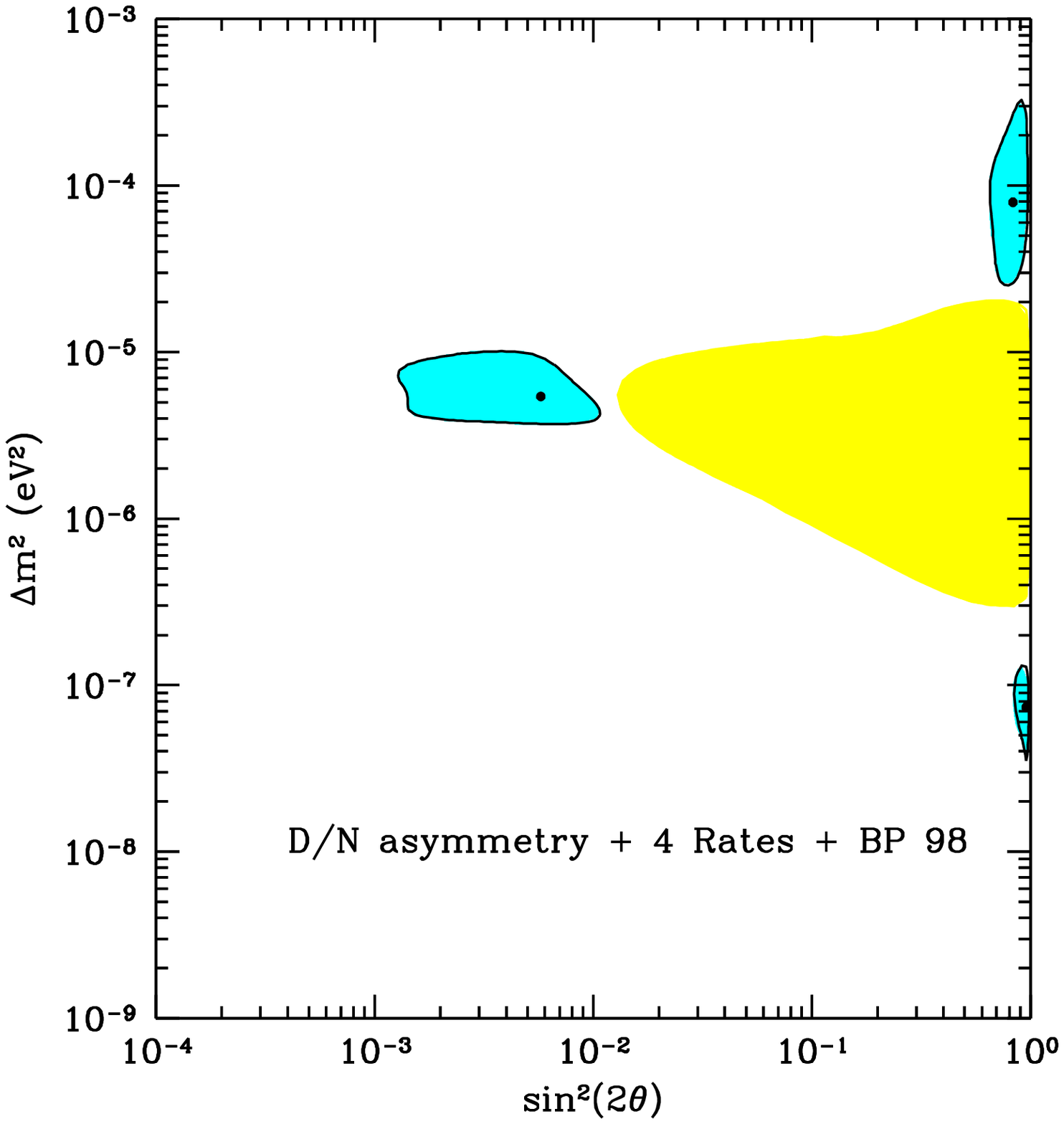}}
\caption[]{ Angular distribution exclusion. 
The region that is excluded by the  SuperKamiokande zenith-angle
 distribution or by the SuperKamiokande Day-Night measurement 
is shown as the light shaded area in the upper and lower 
panels, respectively.  
The darker shaded regions in the upper panel are the regions that are 
allowed by the total measured rates from the chlorine, GALLEX, SAGE, and 
SuperKamiokande experiments plus the zenith angle distribution
measured by SuperKamiokande. The darker shaded regions in the lower
panel are allowed by the four measured rates and the SuperKamiokande
Day-Night asymmetry.}
\label{fig:mswangular}
\end{figure}
}
\begin{figure}
\centerline{\epsfxsize=4in\epsffile{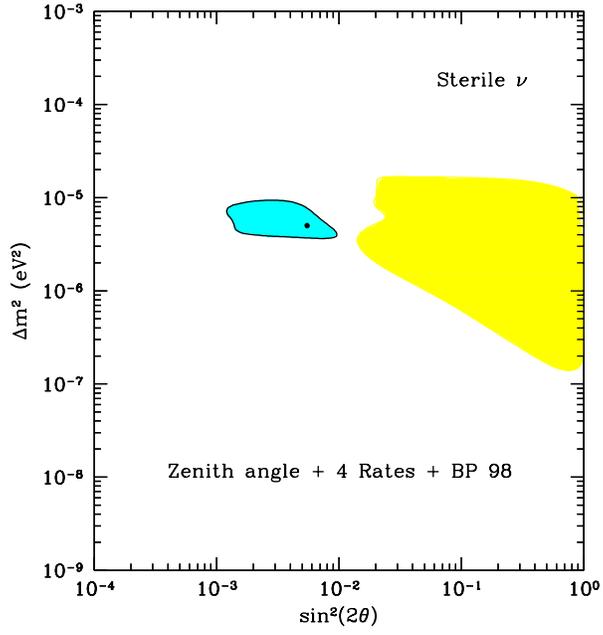}}
\vglue-.5in
\centerline{\epsfxsize=4in\epsffile{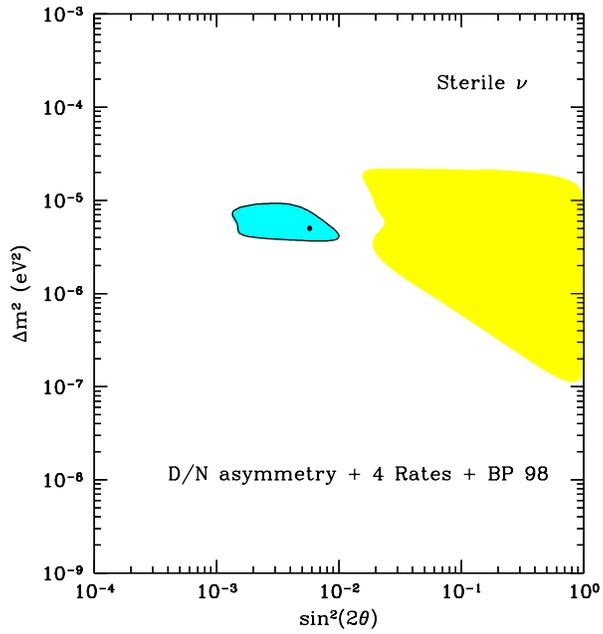}}
\caption[]{Angular distribution exclusion for sterile neutrinos.
This figure is the same as Fig.~\ref{fig:mswangular} except that the
present figure refers to oscillations into sterile neutrinos.}
\label{fig:mswangularsterile}
\end{figure}

\begin{figure} 
\epsfxsize=6.5in\epsffile{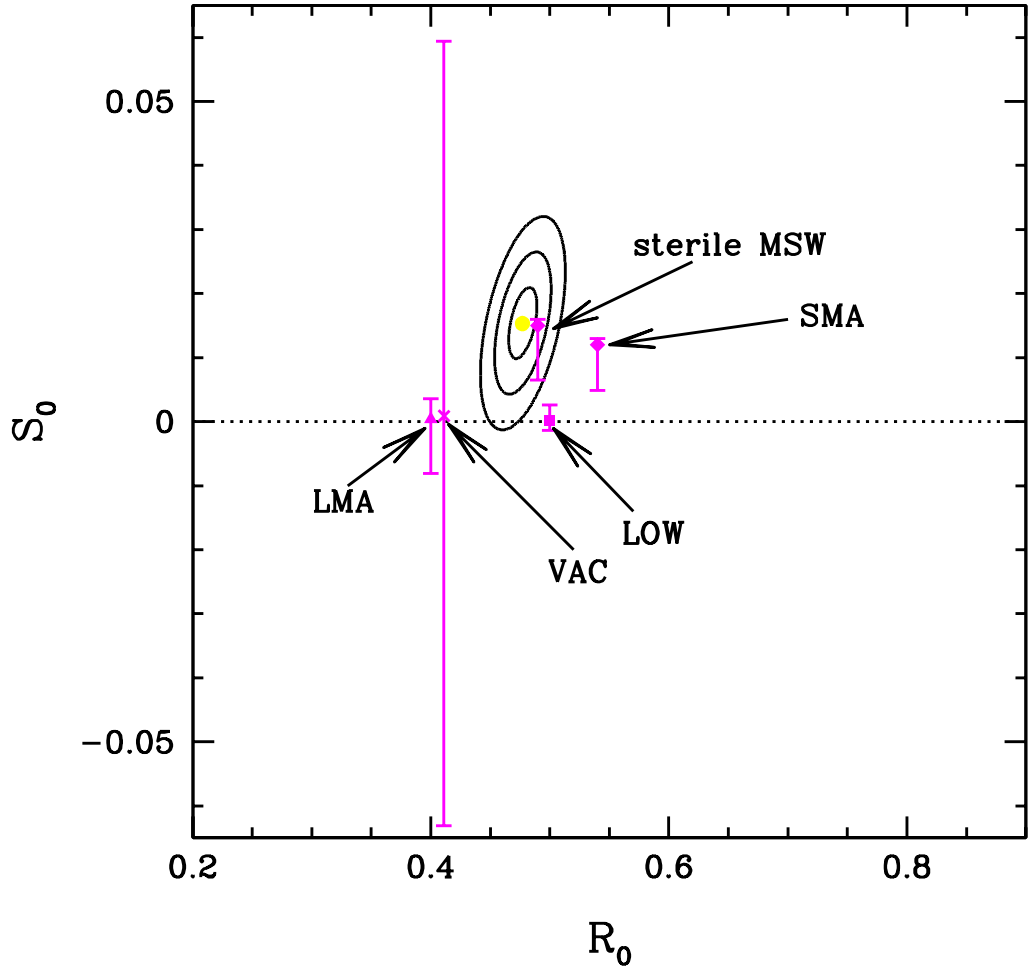}
\caption[]{Deviation from an undistorted energy spectrum.
The $1\sigma$, $2\sigma$, and 
$3\sigma$ allowed regions are shown  in the figure.
The ratio of the observed counting rate as a function of electron 
recoil energy~\cite{superkamiokande504} to the expected
undistorted energy spectrum~\cite{balisi} was fit to a linear function of
energy, with intercept $R_0$ and slope $S_0$ (see
Eq.~\ref{eq:ratioR}).  The five oscillation 
solutions  discussed in Sec.~\ref{averagerates},
SMA active and sterile, LMA, LOW,
and vacuum oscillations,
 all provide
acceptable fits to the data, although the fits are not particularly
good, see text in Sec.~\ref{spectralshape}.}
\label{fig:linefit}
\end{figure}

\vbox{
\begin{figure} 
\centerline{\epsfxsize=3.7in\epsffile{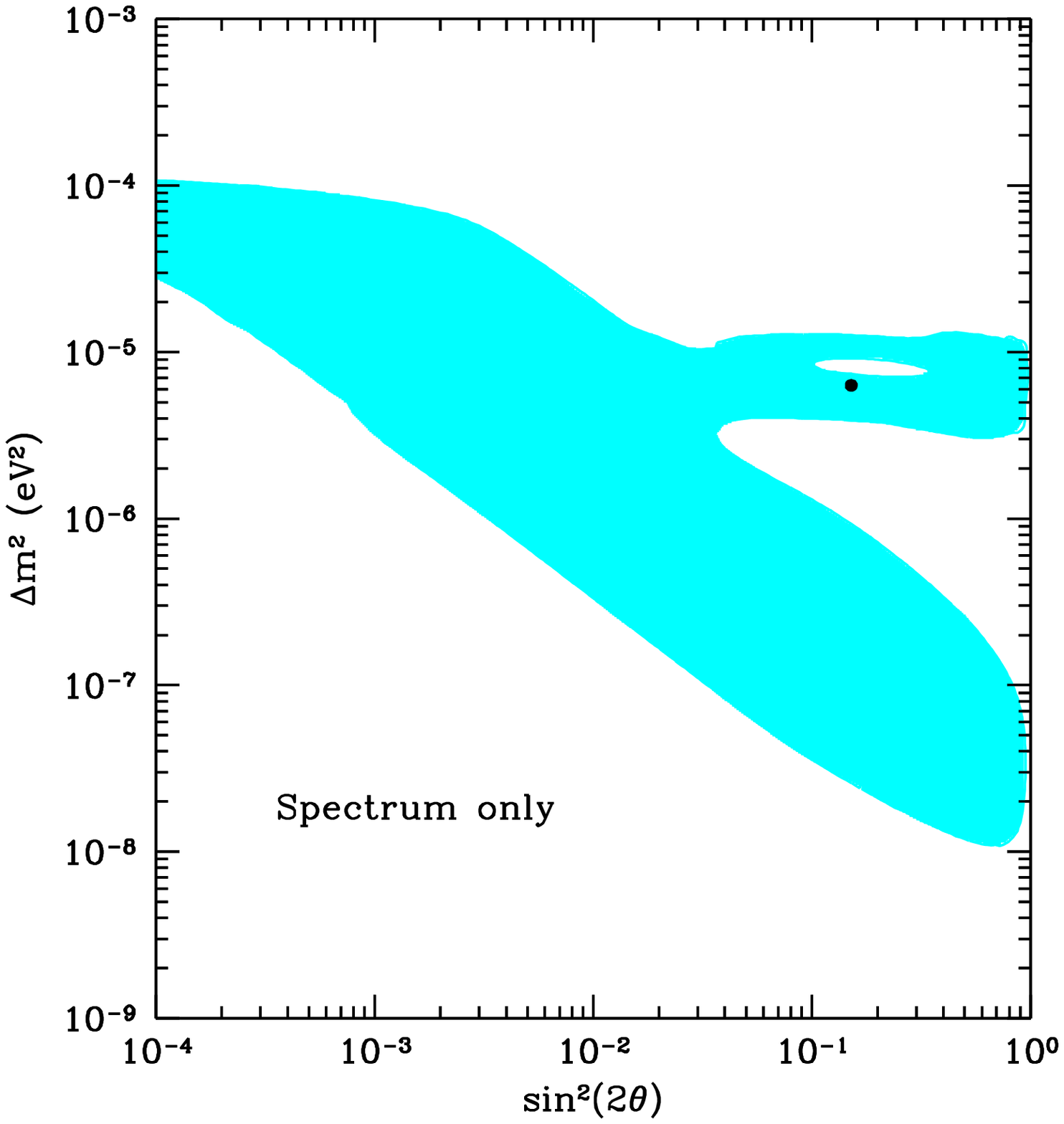}}
\vglue-.6in
\centerline{\epsfxsize=3.7in\epsffile{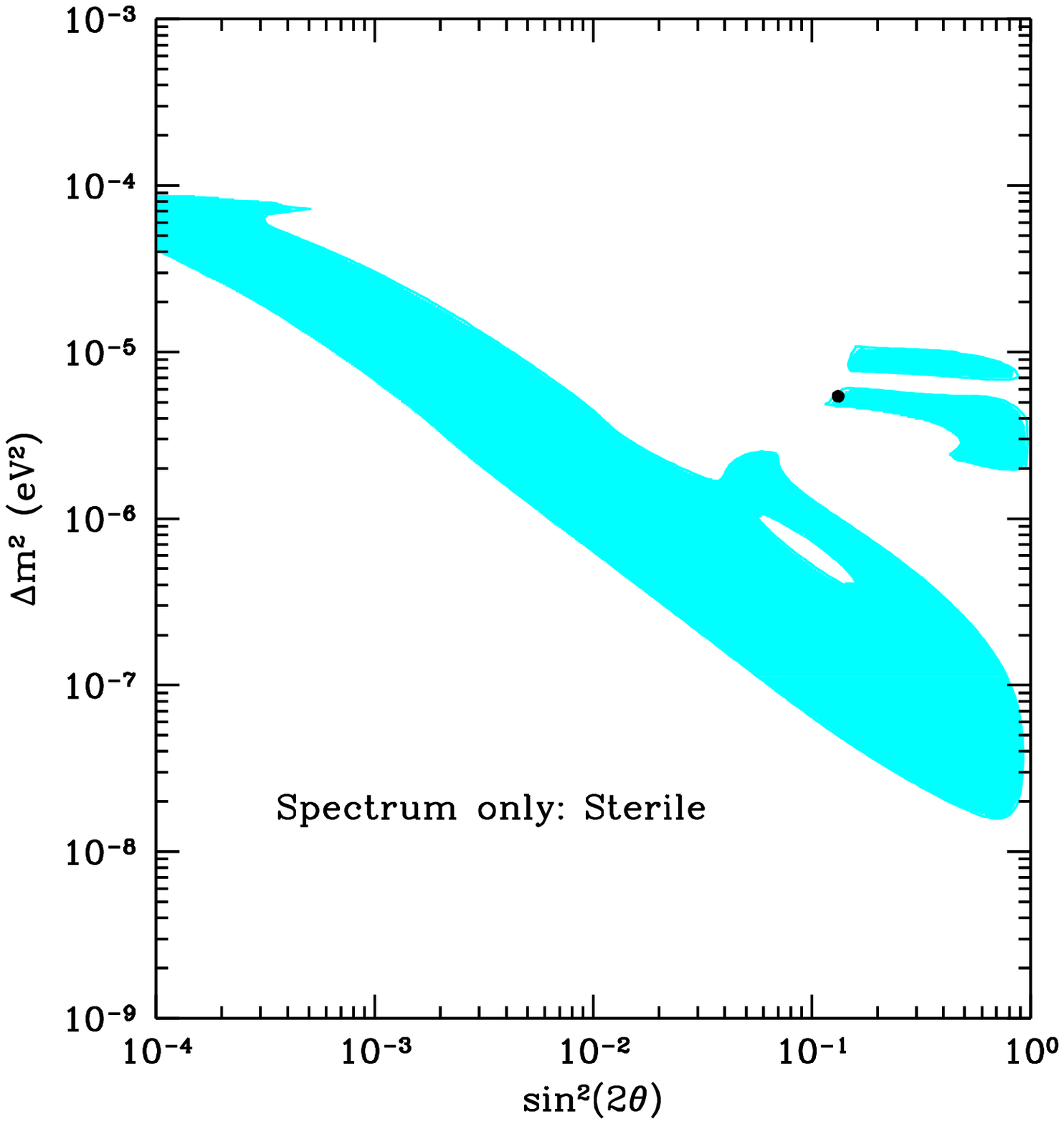}}
\caption[]{Spectrum shape: allowed region  for MSW oscillations.  
Figure~\ref{fig:spectrumMSWa}a refers to active neutrinos and 
Figure~\ref{fig:spectrumMSWa}b refers to sterile neutrinos. 
Each panel 
shows the
region in MSW solution space that is 
allowed  by the SuperKamiokande~\cite{superkamiokande504} 
measurements of the recoil energy spectrum from the scattering of 
${\rm ^8B}$ neutrinos by electrons.  
The best-fit solution when only the energy spectrum is considered is
shown by dark points in Figure~\ref{fig:spectrumMSWa}a and
Figure~\ref{fig:spectrumMSWa}b .
It is interesting to compare 
the regions allowed by the spectral energy distribution with the
regions allowed by the total rates, 
cf. this figure with Fig.~\ref{fig:mswrates} and Fig.~\ref{fig:mswratesst}. }
\label{fig:spectrumMSWa}
\end{figure}
}

\begin{figure} 
\epsfxsize=6.5in\epsffile{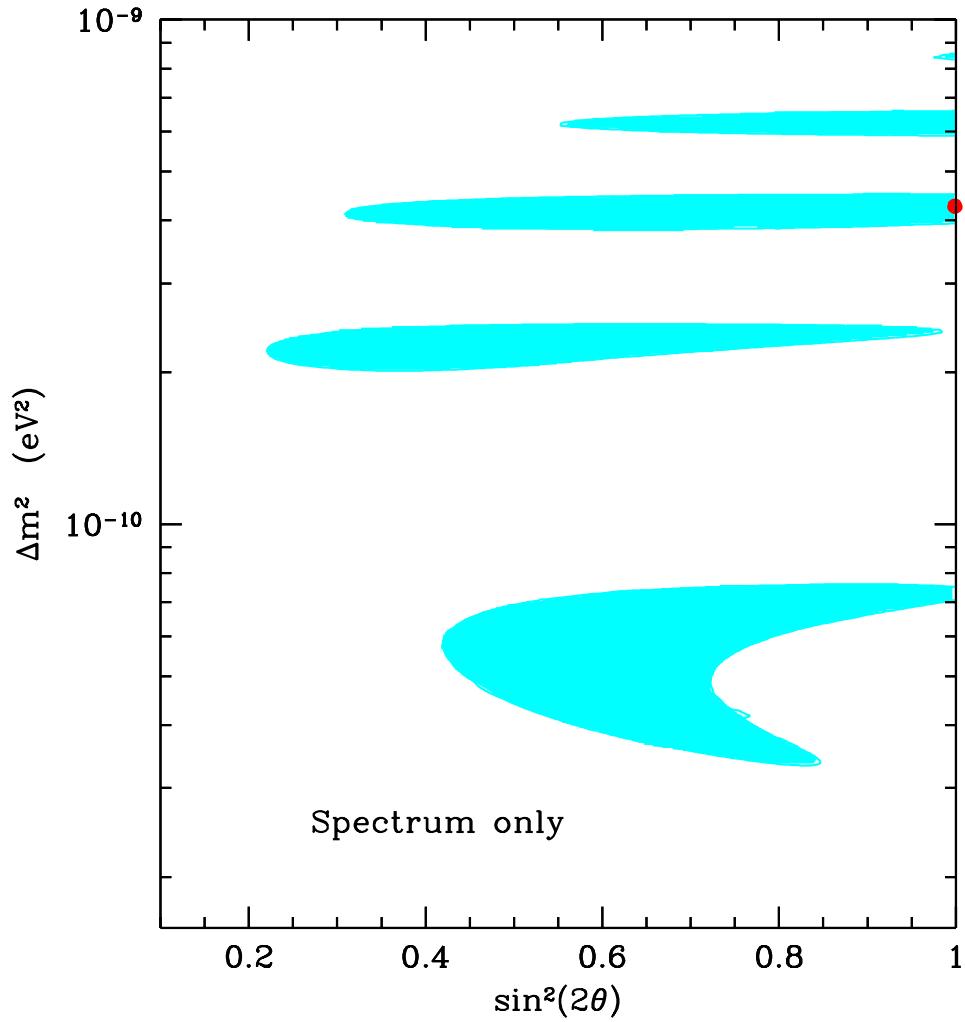}
\caption[]{Spectrum shape: allowed region for  vacuum oscillations.  
The figure shows the allowed 
region in the parameter space of vacuum oscillations  that is 
permitted by the SuperKamiokande~\cite{superkamiokande504} 
measurements of the recoil electron energy spectrum.
cf. Fig.~\ref{fig:vacrates}. The dark point shows the best-fit point
considering the measured spectrum as the only constraint;  the value
of $\chi^2_{\rm min} = 30.5$. }
\label{fig:spectrumV}
\end{figure}

\vbox{
\begin{figure}
\vglue-.5in
\centerline{\epsfxsize=4in\epsffile{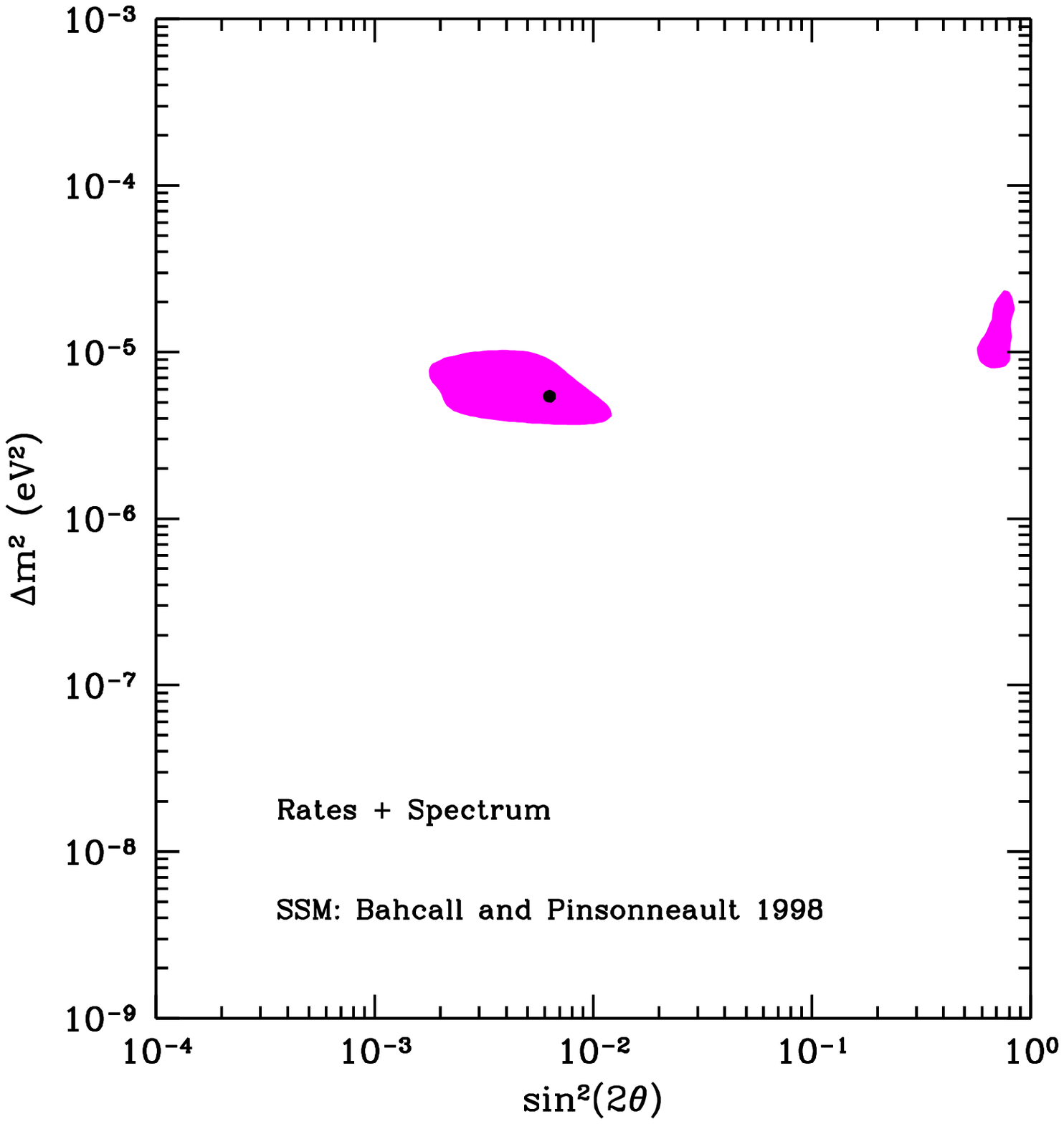}}
\vglue-.5in
\centerline{\epsfxsize=4in\epsffile{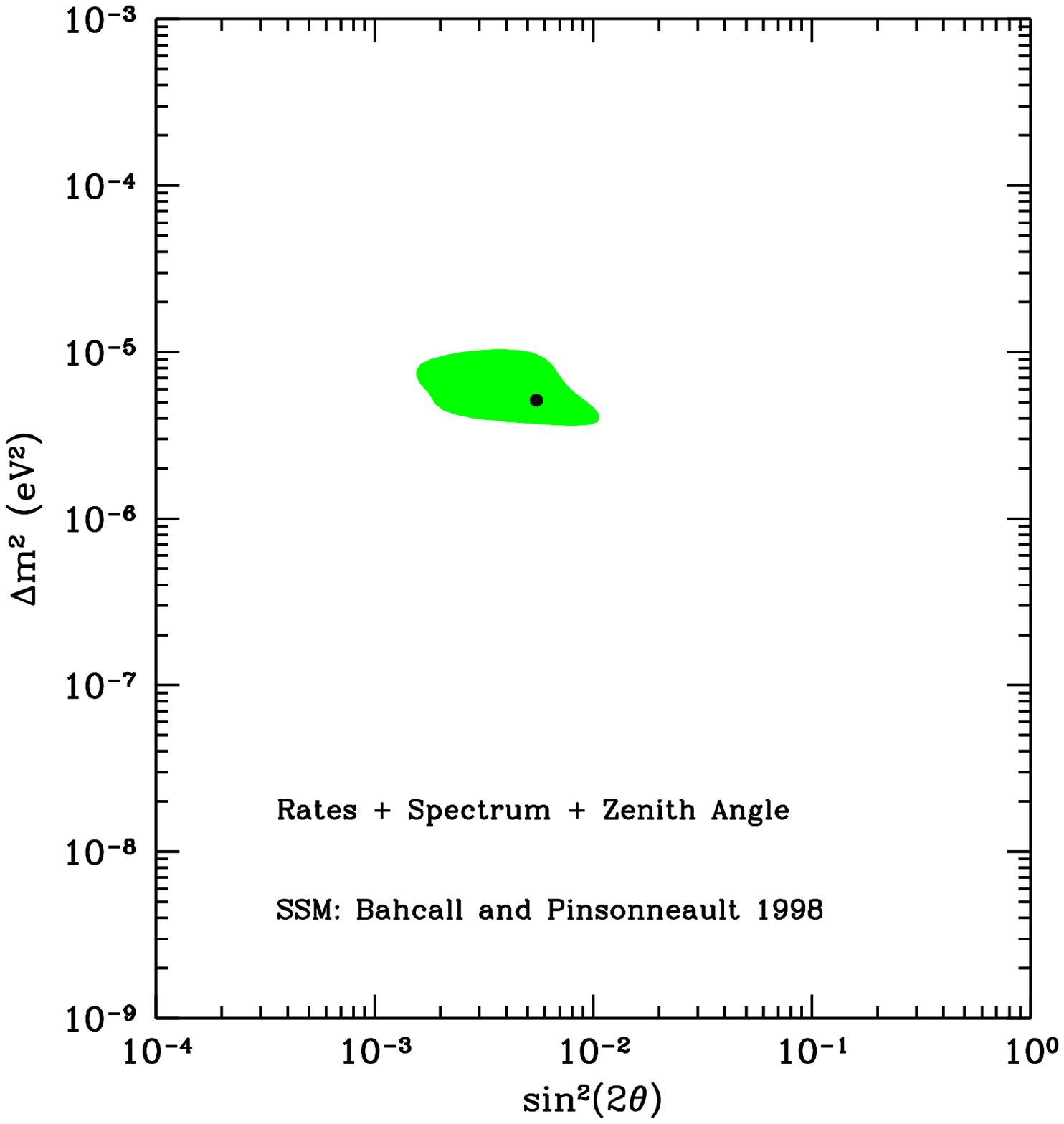}}
\caption[]{Global fits: MSW solutions.
Figure \ref{fig:global}a shows the regions in MSW parameter space that are
consistent with 
the total rates observed in  the four solar neutrino experiments (chlorine,
SuperKamiokande, GALLEX, and SAGE) and the measured SuperKamiokande
electron recoil energy spectrum.  Figure \ref{fig:global}b shows the only
allowed region in MSW parameter space that is consistent with the
combined constraints from the four measured rates and the electron
recoil energy spectrum and zenith angle distribution that are measured
by SuperKamiokande. 
Contours are drawn at $99$\% C.L.
\label{fig:global}}
\end{figure}
}

\begin{figure}
\epsfxsize=6.5in\epsffile{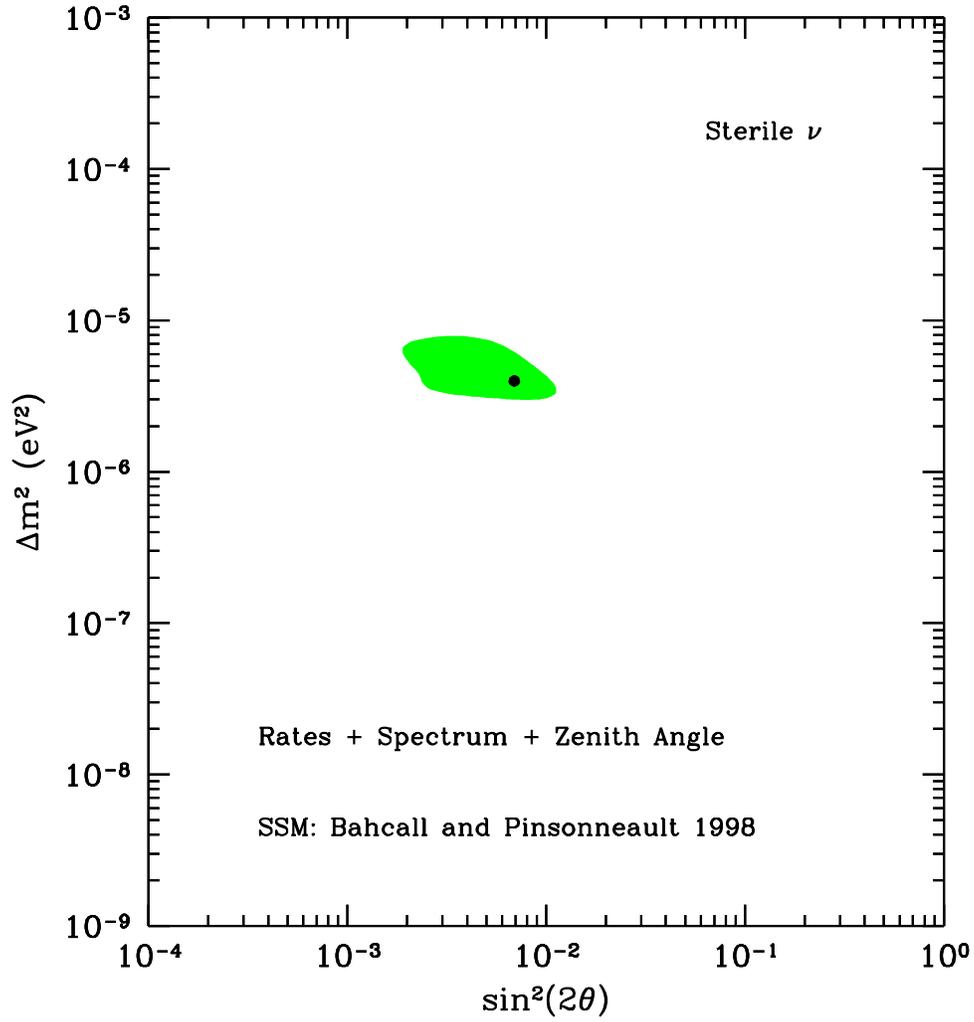}
\caption[]{Global fits: sterile neutrinos. 
The figure shows the allowed parameter
region for MSW oscillations into sterile neutrinos 
that is consistent with the measured
total rates,   the zenith-angle distribution, and 
the  recoil electron energy spectrum.  
Contours are drawn at $99$\% C.L. .}
\label{fig:globalMSWst}
\end{figure}

\begin{figure}
\epsfxsize=6.5in\epsffile{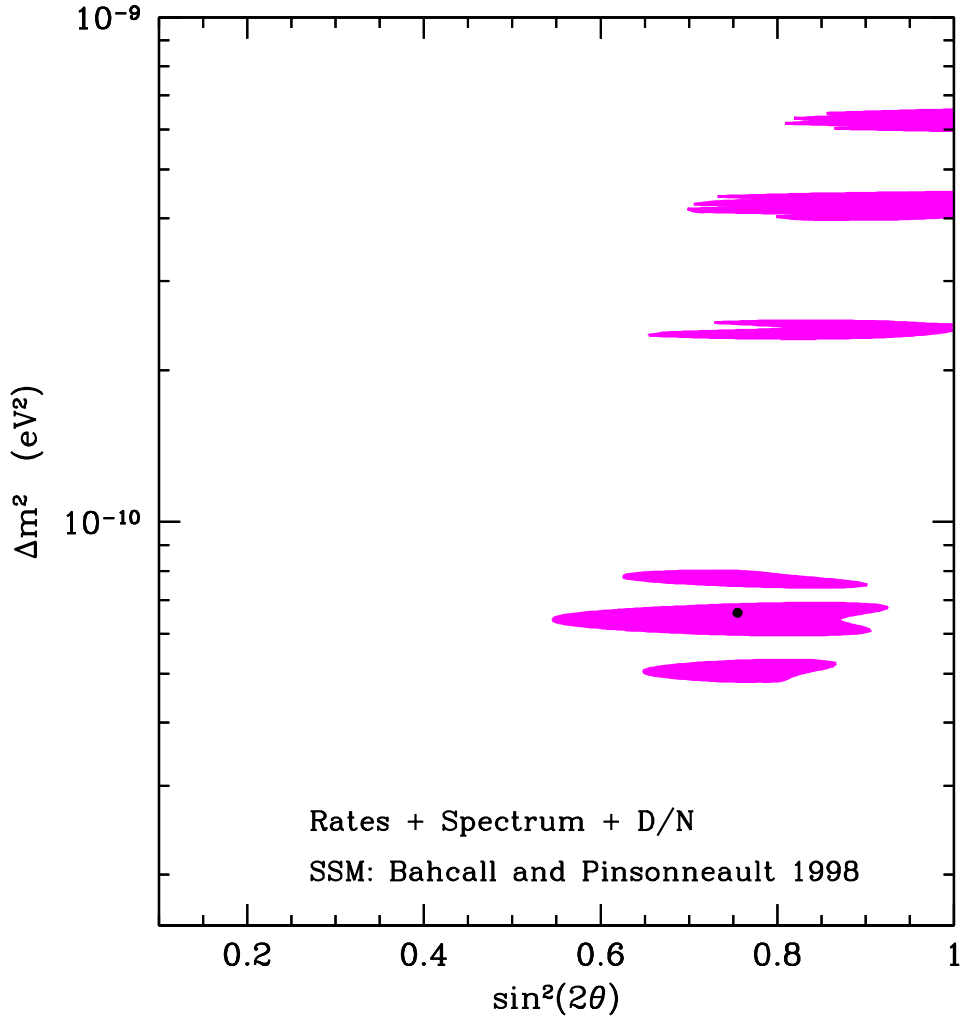}
\caption[]{Global fits: vacuum solutions. 
The figure shows the allowed parameter
region for vacuum oscillations that is consistent with the measured
total rates,  the  recoil electron energy spectrum, 
and the Day-Night asymmetry.  
Contours are drawn at $99$\% C.L. .}
\label{fig:globalV}
\end{figure}

\begin{figure}
\epsfxsize=6.5in\epsffile{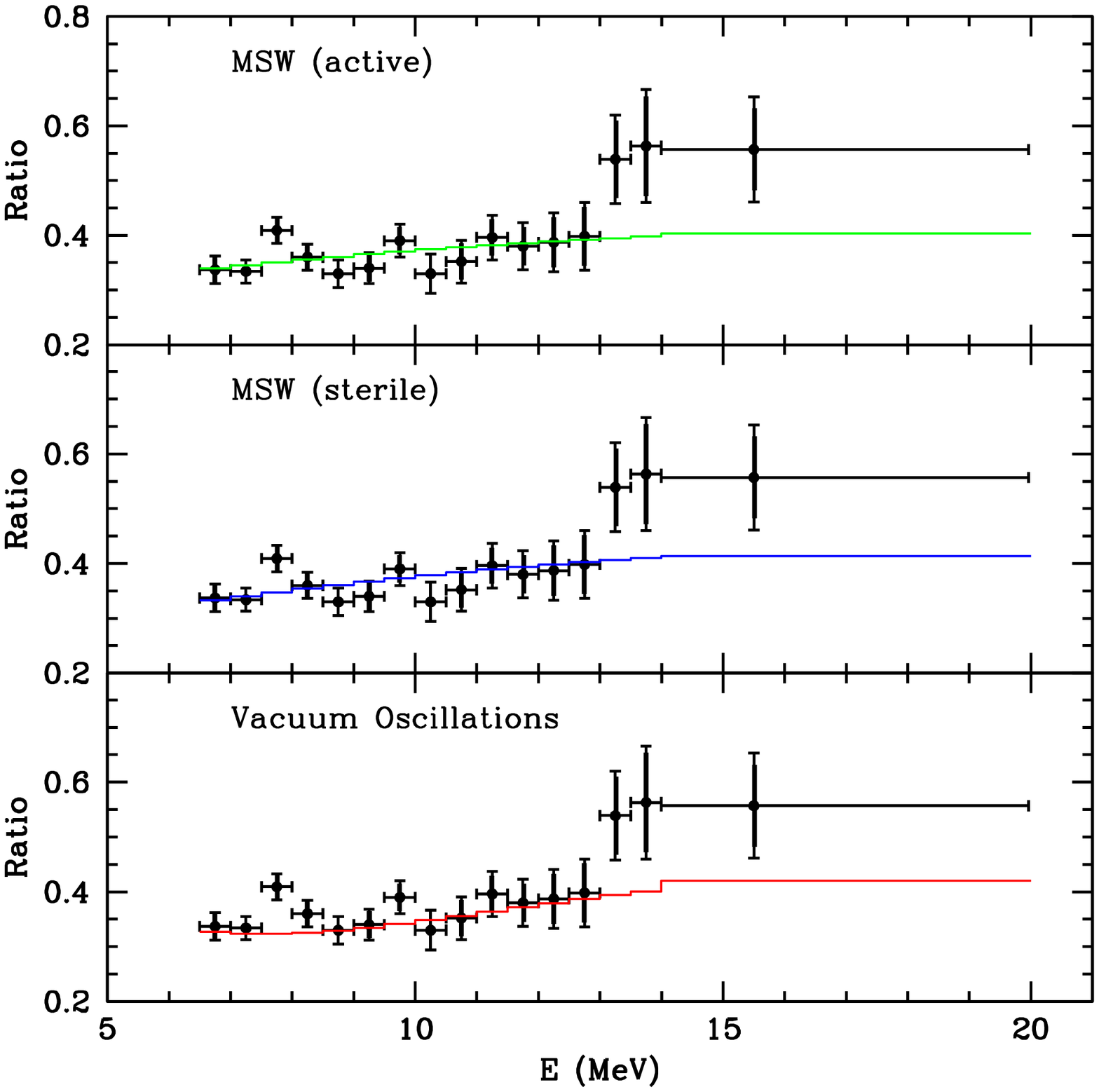}
\caption[]{Global best fits versus measured energy spectrum.
The three panels compare the global neutrino 
oscillation solutions discussed in
Section~\ref{global}  versus the electron
energy spectrum measured by SuperKamiokande and reported at
Neutrino98~\cite{superkamiokande504} . The quantity, Ratio, that is
plotted is the ratio of the number of electrons in a given energy bin,
$E$, 
to the number that is calculated 
using 
the standard, undistorted  $^8$B neutrino
energy spectrum~\cite{balisi} and 
electroweak neutrino-electron scattering cross sections with
radiative corrections~\cite{sirlin}. 
The no oscillation solution is a horizontal line, which,following the
SuperKamiokande collaboration~\cite{superkamiokande504}, 
is normalized to the BP95 prediction~\cite{BP95}
and lies at ${\rm Ratio} = 0.37$.}
\label{fig:globalspectrum}
\end{figure}


\newpage
\begin{thebibliography}{99}
\bibitem{davis68}R. Davis, Jr., D. S. Harmer, and K. C. Hoffman,
Phys. Rev. Lett. {\bf 20}, 1205 (1968).
\bibitem{bahcall68}J. N. Bahcall, N. A. Bahcall, and G. Shaviv,
Phys. Rev. Lett. {\bf 20}, 1209 (1968).
\bibitem{bahcall76}J. N. Bahcall and R. Davis, Jr., Science {\bf 191},
264 (1976).
\bibitem{bahcall89}J. N. Bahcall, {\it Neutrino Astrophysics}
(Cambridge University Press, Cambridge, England, 1989).
\bibitem{kamiokande}KAMIOKANDE Collaboration, Y. Fukuda {\it et al.}, 
Phys. Rev. Lett. {\bf 77}, 1683 (1996).
\bibitem{GALLEX} GALLEX Collaboration, P.~Anselmann {\it et al.}, 
Phys.~Lett. B {\bf 342}, 440 (1995); GALLEX Collaboration,
W. Hampel {\it et al.},  Phys.~Lett. B 
{\bf 388}, 364 (1996).
\bibitem{SAGE}SAGE Collaboration, V. Gavrin {\it et al.}, in {\it
Neutrino 96}, Proceedings of 
the XVII International
Conference on Neutrino Physics and Astrophysics, Helsinki, edited by  
K. Huitu, K. Enqvist and J. Maalampi 
(World Scientific, Singapore, 1997), p. 14;
V. Gavrin {\it et al.}, in {\it Neutrino 98}, Proceedings of the XVIII International Conference on
Neutrino Physics and Astrophysics, Takayama, Japan, 4--9 June 1998,
edited by Y. Suzuki and Y. Totsuka. To be published in Nucl. Phys. B
(Proc. Suppl.).
\bibitem{superkamiokande300} SuperKamiokande Collaboration, Y. Fukuda {\it
et al.} Phys. Rev. Lett. (accepted for publication, 
hep-ex/9805021).
\bibitem{superkamiokande374}R.~Svoboda~(for~the~SuperKamiokande~Collaboration),
ITP Conference on Solar neutrinos: News About SNUs, December 2--6,
1997 (unpublished), http://www.itp.ucsb.edu/online/snu/svoboda/; 
H. Sobel (for the
SuperKamiokande Collaboration), Aspen Winter Conference on Particle
Physics, January 25-31, 1998, Aspen, CO (unpublished).
http://www.physics.arizona.edu/~ina/aspen98.html.
\bibitem{superkamiokande504} SuperKamiokande Collaboration, Y. Suzuki, in 
{\it Neutrino 98}, Proceedings of the XVIII International Conference on
Neutrino Physics and Astrophysics, Takayama, Japan, 4--9 June 1998,
edited by Y. Suzuki and Y. Totsuka. To be published in Nucl. Phys. B
(Proc. Suppl.). 
\bibitem{ref:msw}L. Wolfenstein, Phys. Rev. D {\bf 17}, 2369 (1978);
S. P. Mikheyev and A. Yu. Smirnov, Yad. Fiz. {\bf 42}, 1441 (1985)
[Sov. J. Nucl. Phys. {\bf 42}, 913 (1985)]; Nuovo 
Cimento C {\bf 9}, 17 (1986).
\bibitem{ref:vac}V. N. Gribov and B. M. Pontecorvo, Phys. Lett. B {\bf
28}, 493 (1969).
\bibitem{McD94}A. B.~McDonald, in {\it Proceedings of the 9th Lake 
Louise Winter Institute}, edited by A.~Astbury {\it et al.} 
(World Scientific, 1994), p. 1.
\bibitem{borexino}C.~Arpesella et al., BOREXINO proposal, Vols. 1
and 2, eds. G.~Bellini, R.~Raghavan et al. (Univ. of
Milano, Milano, 1992).
\bibitem{bahcallga97}J. N. Bahcall, Phys. Rev. C {\bf 56}, 3391 (1997).
\bibitem{bahcalletal96}J. N. Bahcall {\it et al.}, Phys. Rev. C {\bf 54},
411 (1996). 
\bibitem{bahcall88}J. N. Bahcall and R. Ulrich, Rev. Mod. Phys. {\bf
60}, 297 (1988).
\bibitem{sirlin}J.~N.~Bahcall, M.~Kamionkowski, and A.~Sirlin, 
Phys.\ Rev.\ D {\bf 51}, 6146 (1995). 
\bibitem{balisi}  J.~N.~Bahcall, E.~Lisi, D.~E.~Alburger, L.~De Braeckeleer,
S.~J.~Freedman, and J.~Napolitano, Phys.\ Rev.\ C {\bf  54}, 411 (1996).
\bibitem{BP98}J. N. Bahcall, S. Basu, and M. H. Pinsonneault,
 to be published in Phys. Lett. B, July 1998, astro-ph/9805135.
\bibitem{adelberger98}E. Adelberger {\it et al.},
Rev. Mod. Phys. (to be published, October 1998).
\bibitem{johnson92}C. W. Johnson, E. Kolbe, S. E. Koonin, 
and K. Langanke, Astrophys. J. {\bf 392}, 320 (1992).
\bibitem{giunti}S. M. Bilenky and C. Giunti, ``Implications of CHOOZ
Results for the Decoupling of Solar and Atmospheric Neutrino 
Oscillations,'' hep-ph/9802201. 
\bibitem{chooz}CHOOZ experiment: M. Apollonio {\it et al.},
Phys. Lett. B {\bf 420}, 397 (1998).
\bibitem{reliable}J. N. Bahcall, M. H. Pinsonneault, S. Basu, and
J. Christensen-Dalsgaard, Phys. Rev. Lett. {\bf 78}, 171 (1997).
\bibitem{BP92} J. N. Bahcall and M. H. Pinsonneault, Rev. Mod. Phys.
{\bf 64}, 885 (1992).
\bibitem{BP95}J. N. Bahcall and M. H. Pinsonneault, Rev. Mod. Phys.
{\bf 67}, 781 (1995).
\bibitem{hata94}N. Hata, S. Bludman, and P. Langacker, Phys Rev D {\bf
49}, 3622 (1994).
\bibitem{parke95}S. Parke, Phys. Rev. Lett. {\bf 74}, 839 (1995).
\bibitem{robertson}K. M. Heeger and R. G. H. Robertson, Phys Rev Lett
{\bf 77}, 3720 (1996).
\bibitem{howwell}J. N. Bahcall and P. I. Krastev, Phys. Rev. D {\bf
53}, 4211 (1996).
\bibitem{bethe90}J. N. Bahcall and H. A. Bethe, Phys. Rev. Lett. {\bf
65}, 2233 (1990).
\bibitem{hata95} N. Hata and P. Langacker, Phys. Rev. D {\bf 52}, 
420 (1995).
\bibitem{fogli95}G. L. Fogli, E. Lisi, and D. Montanino, Phys. Rev. D
{\bf 49}, 3226 (1994); see also G. L. Fogli and E. Lisi, Astropart.
Phys. {\bf 3}, 185 (1995).
\bibitem{berezinsky95} V. Berezinsky, G. Fiorentini, and M. 
Lissia, Phys. Lett. B {\bf 341}, 38 (1995). 
\bibitem{HL} N. Hata and P. Langacker, Phys. Rev. D {\bf 50}, 632
(1993).
\bibitem{KPanal} P. Krastev and S. T. Petcov, Phys. Rev. Lett. {\bf
72}, 1960 (1994);  Nucl. Phys. B {\bf 449}, 605 (1995).
\bibitem{Krauss} E. Gates, L. M. Krauss, and M. White, 
Phys. Rev. D {\bf 51}, 2631 (1995).
\bibitem{Kuo} T. K. Kuo and J. Pantaleone, Rev. Mod. Phys. 
{\bf 61}, 937 (1987).
\bibitem{fio} E. Calabresu, N. Ferrari, G. Fiorentini, and M. Lissia, 
 preprint INFN-FE-10-95, June 1995, hep-ph/9507352 (unpublished).
\bibitem{maris97} M. Maris and S. T. Petcov, Phys. Rev. D {\bf 56}, 7444
(1997). 
\bibitem{rossi} Z. Berezhiani and A. Rossi, Phys. Rev. D {\bf 51},
5229 (1995).
\bibitem{brighter}J. N. Bahcall and P. I. Krastev, Phys. Rev. C {\bf 56},
2839 (1997).
\bibitem{KP88} P. Krastev and S. T. Petcov, Phys. Lett. B {\bf 207}, 64  
(1988); [Erratum: B214, 661 (1988)]. See also S. T. Petcov, Phys.
Lett. B {\bf 200}, 373 (1988); {\bf 214}, 139 (1988).
\bibitem{homestake}B. T. Cleveland {\it et al.}, Astrophys. J. {\bf
496}, 505 (1998); B. T. Cleveland {\it et al.}, 
Nucl. Phys. B (Proc. Suppl.) {\bf 38}, 47 (1995); R. Davis,
Prog. Part. Nucl. Phys. {\bf 32}, 13 (1994).
\bibitem{krastev93} P. I. Krastev and S. T. Petcov, Phys. Lett. B {\bf
299}, 99 (1993); G. L. Fogli, E. Lisi, and D. Montanino,
Phys. Rev. D {\bf 54}, 2048 (1996).
\bibitem{BaltzW3} A. J. Baltz and J. Weneser, Phys. Rev. D {\bf 50}, 5971
(1994); {\bf 51}, 3960 (1995).
\bibitem{krastevsmirnov94}P. Krastev and A. Smirnov, Phys. Lett. B {\bf
338}, 282 (1994).
\bibitem{krastevpetcov96}P. Krastev and S. Petcov, Phys. Rev. D {\bf
53}, 1665 (1996).
\bibitem{nussinov76}S. Nussinov, Phys. Lett. B {\bf 63}, 201 (1976).
\bibitem{harrison}P. F. Harrison, D. H. Perkins, and W. G. Scott,
Phys. Lett. B {\bf 349}, 137 (1995); {\bf 396}, 186 (1997).
\bibitem{foot}R. Foot and R. R. Volkas, Phys. Rev. D {\bf 52}, 6595 (1995).
\bibitem{comforto}G. Comforto, A. Marchionni, F. N. Martelli,
F. Vetrano, M. Lanfranchi, and G. Torricelli-Ciamponi,
Astropart. Phys. {\bf 5}, 147 (1996).
\bibitem{acker}A. Acker and S. Pakvasa, Phys. Lett. B {\bf 397}, 209
(1997); A. Acker, S. Pakvasa, J. Leaned and T. Weiler, 
Phys. Lett. B {\bf  298}, 149 (1993).
\bibitem{krastevpetcov97}P. I. Krastev and S. T. Petcov, Phys. Lett. B
{\bf 395}, 69 (1997).
\bibitem{earthreg} S. P. Mikheyev and A. Yu. Smirnov, in {\it '86
Massive Neutrinos in Astrophysics and in Particle Physics},
Proceedings of the Sixth Moriond Workshop, edited by O. Fackler and
Y. Tr\^an Thanh V\^an (Editions Fronti\`eres, Gif-sur-Yvette, 1986),
p. 355;
J. Bouchez {\it et al.}, Z. Phys. C {\bf 32}, 499 (1986); M. Cribier,
W. Hampel, J. Rich, and D. Vignaud, Phys. Lett. B {\bf 182}, 89 (1986);
M. L. Cherry and K. Lande, Phys. Rev. D {\bf 36}, 3571 (1987);
S. Hiroi, H. Sakuma, T. Yanagida, and M. Yoshimura, Phys. Lett. B {\bf
198}, 403 (1987); S. Hiroi, H. Sakuma, T. Yanagida, and M. Yoshimura,
Prog. Theor. Phys. {\bf 78}, 1428 (1987); A. Dar, A. Mann, Y. Melina,  
and D. Zajfman, Phys. Rev. D {\bf 35}, 3607
(1988); M. Spiro
and D. Vignaud, Phys. Lett. B {\bf 242}, 279 (1990).
\bibitem{BaltzW} A. J. Baltz and J. Weneser, Phys. Rev. D {\bf 35}, 528 (1987).
\bibitem{BaltzW2} A. J. Baltz and J. Weneser, Phys. Rev. D {\bf 37}, 3364 
(1988).
\bibitem{bahcall91}J. N.~Bahcall, Phys.~Rev. D {\bf 44}, 1644
(1991).
\bibitem{refstraightline}W. Kwong and S. P. Rosen,
Phys. Rev. Lett. {\bf 68}, 748 (1992); 
P. I. Krastev and A. Yu. Smirnov, Phys. Lett. B {\bf 338}, 282 (1994);
W. Kwong and S. P. Rosen, 
Phys. Rev. D {\bf 44}, 2241 (1995);
P. I. Krastev and S. T. Petcov, Nucl. Phys. B {\bf 449}, 605 (1995).
\bibitem{raghavan97}R. S. Raghavan, Phys. Rev. Lett. {\bf 78}, 3618 (1997). 
\bibitem{hep98} J. N. Bahcall and P. Krastev, Phys. Lett. B (in press)
(1998), hep-ph/9807525.
\bibitem{gno}E. Bellotti {\it et al.}, ``Proposal for a Permanent Gallium
Neutrino Observatory (GNO) at Laboratori Nazionali Del Gran Sasso,''
http://kosmopc.mpi-hd.mpg.de/gallex/gallex.html, February (1996).
\bibitem{models}(GONG) J. Christensen-Dalsgaard {\it et al.}, GONG
Collaboration, Science {\bf 272}, 1286 (1996); 
 (BP95) J. N. Bahcall and M. H. Pinsonneault, Rev. Mod. Phys.
{\bf 67}, 781 (1995); (KS94) A. Kovetz and G. Shaviv, Astrophys. J. {\bf
426}, 787 (1994); (CDF94) V. Castellani, S. Degl'Innocenti, G. Fiorentini,
L. M. Lissia, and B. Ricci, Phys. Lett. B {\bf 324}, 425 (1994);
(JCD94) J. Christensen-Dalsgaard, Europhys. News {\bf 25}, 71 
(1994); (SSD94) X. Shi, 
D. N. Schramm, and D. S. P. Dearborn, Phys. Rev. D {\bf 50}, 2414
(1994); (DS96) A. Dar and G. Shaviv, Astrophys. J. {\bf 468}, 933
(1996); (CDF93) V. Castellani, S. Degl'Innocenti, and G. Fiorentini, Astron. 
Astrophys. {\bf 271}, 601 (1993); (TCL93) S. Turck-Chi\`eze and I. Lopes, 
Astrophys. J. {\bf 408}, 347 (1993); (BPML93) G. Berthomieu, J. Provost, P. 
Morel, and Y. Lebreton, Astron. Astrophys. {\bf 268}, 775 (1993);
(BP92) J. N. Bahcall and M. H. Pinsonneault, Rev. Mod. Phys. {\bf 64}, 885
(1992); (SBF90) I.-J. Sackman, A. I. Boothroyd, and W. A. Fowler,
Astrophys. J. {\bf 360}, 727 (1990); (BU88) J. N. Bahcall and R. K. Ulrich, 
Rev. Mod. Phys. {\bf 60}, 297 (1988); (RVCD96) O. Richard,
S. Vauclair, C. Charbonnel, and W. A. Dziembowski,
Astron. Astrophys. {\bf 312}, 1000 (1996); (CDR97) F. Ciacio,
S. Degl'Innocenti, and B. Ricci, Astron. Astrophys. Suppl. Ser. 
{\bf 123}, 449 (1997).

\end{thebibliography}
\end{document}